\documentclass[10pt]{article}
\usepackage[usenames,dvipsnames]{color}
\usepackage{amsmath,amsthm,amsfonts,amssymb,multicol,amscd,amsbsy}
\usepackage{graphicx}
\usepackage{booktabs}
\usepackage{array}
\usepackage{subfigure}
\usepackage{enumerate}
\usepackage{url}
\usepackage[nodayofweek]{datetime}
\usepackage[english]{babel}
\usepackage[toc,page]{appendix}
\usepackage{verbatim} 
\usepackage{amsthm} 
\usepackage{enumitem}
\usepackage{enumerate}
\usepackage{bbm} 
\usepackage[normalem]{ulem} 
\usepackage{bm}
\usepackage{hyperref}

\usepackage[top=2.75cm,bottom=3.75cm]{geometry}

\usepackage{authblk}

\newcommand{\be}{\begin}
\newcommand{\e}{\end}
\newcommand{\beq}{\begin{equation}}
\newcommand{\eeq}{\end{equation}}
\newcommand{\beqs}{\begin{equation*}}
\newcommand{\eeqs}{\end{equation*}}
\newcommand{\bal}{\begin{align}}
\newcommand{\eal}{\end{align}}
\newcommand{\bals}{\begin{align*}}
\newcommand{\eals}{\end{align*}}
\newcommand{\nn}{\nonumber}

\newcommand{\T}{\mathbb{T}}
\newcommand{\N}{\mathbb{N}}
\newcommand{\Q}{\mathbb{Q}}

\renewcommand{\Im}{\mathrm{Im}}

\newcommand{\set}[1]{\mathbb{#1}}

\newcommand{\R}{\set{R}}
\newcommand{\C}{\set{C}}
\newcommand{\Z}{\set{Z}}

\newcommand{\lv}{\left\lVert}
\newcommand{\rv}{\right\rVert}

\newtheorem{thm}{Theorem}[section]
\newtheorem{lm}[thm]{Lemma}
\newtheorem{cor}[thm]{Corollary}
\newtheorem{prop}[thm]{Proposition}

\theoremstyle{definition}

\numberwithin{equation}{section}

\theoremstyle{remark}
\newtheorem{rmk}[thm]{Remark}

\def\dotuline{\bgroup
  \ifdim\ULdepth=\maxdimen  
   \settodepth\ULdepth{(j}\advance\ULdepth.4pt\fi
  \markoverwith{\begingroup
  \advance\ULdepth0.08ex
  \lower\ULdepth\hbox{\kern.15em .\kern.1em}%
  \endgroup}\ULon}

\def\dashuline{\bgroup
  \ifdim\ULdepth=\maxdimen  
   \settodepth\ULdepth{(j}\advance\ULdepth.4pt\fi
  \markoverwith{\kern.15em
  \vtop{\kern\ULdepth \hrule width .3em}%
  \kern.15em}\ULon}
\allowdisplaybreaks

\begin{document}
\setlength{\columnsep}{5pt}
\title{Large deviation estimates  and H\"older regularity of the Lyapunov exponents for quasi-periodic Schr\"odinger cocycles}

\author{Rui Han and Shiwen Zhang}


\date{}
\maketitle
\begin{abstract}
We consider one-dimensional quasi-periodic Schr\"odinger operators with analytic potentials.
In the positive Lyapunov exponent regime, we prove large deviation estimates which lead to refined H\"older continuity of the Lyapunov exponents and the integrated density of states,
in both small Lyapunov exponent and large coupling regimes. Our results cover all the Diophantine frequencies and some Liouville frequencies.
\end{abstract}

\section{Introduction and the Main Results}
In this paper, we study the following one dimensional discrete quasi-periodic operators on $\ell^2(\Z)$:
\begin{equation}\label{def:Schop}
  (H(x) \varphi)(n)=\varphi(n-1)+\varphi(n+1)+v(x+n\omega)\varphi(n),\ n\in\Z,
\end{equation}
where $x\in \T:=[0,1]$ is called phase, $\omega\in\T\setminus \Q$ is called frequency and the real valued analytic function $v:\T\to\R$ is called potential. 

For an energy $E\in\R$, the Schr\"odinger equation
\beq\label{def:Scheq}
\varphi(n-1)+\varphi(n+1)+v(x+n\omega)\varphi(n)=E\varphi(n)
\eeq
can be rewritten in the form of the skew-product:
\beq\label{eq:transfernton+1}
 \left( \begin{array}{c}
    \varphi(n+1) \\
    \varphi(n)
  \end{array}\right)=
  A(\omega,E; x+n\omega)
  \left( \begin{array}{c}
    \varphi(n) \\
    \varphi(n-1)
  \end{array}\right),
\eeq
  where 
  \beq\label{def:A}
  A(\omega,E; x):=
  \left(\begin{array}{cc}
E- v(x)& -1 \\
1 & 0
\end{array}\right).
  \eeq
  The dynamical system $(\omega, A(\omega, E; \cdot)): \T\times \C^2 \to \T\times \C^2$, defined by
  \beq
(\omega, A(\omega, E; \cdot ))(x, v)=(x+\omega, A(\omega, E; x)v),  
  \eeq
  is called Schr\"odinger cocycle.
  
  Let $A$ be defined as in \eqref{def:A} and let 
  	\beq\label{def:M}
  	M_n(\omega,E; x):=A(\omega,E; x+n\omega)A(\omega,E; x+(n-1)\omega)\cdots A(\omega, E; x+\omega),
  	\eeq 
  be the $n$-step transfer matrix, coming from $n$ iterates of the Schr\"odinger cocycle $(\omega, A)$.
  Then, in view of \eqref{eq:transfernton+1}, one clearly has
\begin{equation}\label{eq:transfer0ton}
 \left( \begin{array}{c}
    \varphi(n+1) \\
    \varphi(n)
  \end{array}\right)=
  M_n(\omega, E; x)
  \left( \begin{array}{c}
    \varphi(1) \\
    \varphi(0)
  \end{array}\right).
\end{equation}
Let
\begin{equation}\label{def:unLn}
  u_n(\omega, E;x):=\frac{1}{n}\log\| M_n(\omega,E;x)\|,\ \
  \text{and}\ \ 
  L_n(\omega,E):=\int_{\T}u_n(\omega, E; x){\rm d} x.
\end{equation}
For any irrational $\omega\in [0,1]$, the translation $x\mapsto x+\omega$ is ergodic. 
The Furstenberg-Kesten theorem implies that the following limit exists for a.e. $x$:
\beq\label{def:Lyp}
\lim_{n\rightarrow \infty} u_n(\omega, E; x)=\lim_{n\rightarrow\infty} L_n(\omega, E)=:L(\omega, E).
\eeq
The limit $L(\omega,E)$ is called the \emph{Lyapunov exponent}.  
Let us point out that in the definition of quasi-perodic cocycle, one could in general replace the one dimensional rotation number $\omega\in \T$ by a higher dimensional vector $\omega\in \T^d$, and could also replace $A$ by any $m\times m$ matrix-valued function, where $m\in \N$.
The definition \eqref{def:Lyp} then yields the maximal Lyapunov exponent $L(\omega, A)$.

Note that for any fixed $\kappa>0, E, \omega$, the a.e. convergence in (\ref{def:Lyp}) implies
\begin{equation}\label{eq:LDTrough}
  {\rm mes}\big\{x\in\T:|u_n(\omega,E;x)-L_n(\omega,E)|> \kappa\big\}\to0\ \ {\rm as}\ \ n\to\infty.
\end{equation}
Thus the question lies in the convergence rate w.r.t. $n$ and the dependence on $\omega$, $E$ and $v$. 
Such estimate is in general known as the \emph{Large Deviation Theorem/Principle(LDT/LDP)} in probability theory. In this paper, we shall focus on the LDT for the monodromy matricies as introduced in (\ref{def:unLn}). 
For the general LDT theory in probability theory, we refer readers to \cite{RSbook,DKbook}. 

Another important quantity in the spectral theory of Schr\"odinger operators is the
\emph{integrated density of states} (I.D.S.), denoted by $N$. 
It is also a function of the energy $E$.
The I.D.S. gives the asymptotic distribution of eigenvalues of $H$ restricted to large boxes. 
It is linked to $L(E)=L(\omega,E)$ via the Thouless formula, see e.g. \cite{CS},
$$L(E)=\int\log|E-E{'}|{\rm d}N(E{'}).$$
The I.D.S. is in general continuous in $E$, but this does not directly imply the continuity of the Lyapunov exponent.
However by virtue of the Hilbert-transform, H\"older regularities of $N(E)$ and $L(E)$ pass from one to the other. 
For a proof of this fact, see e.g. \cite{GS1}.
Therefore we shall focus on the Lyapunov exponent in the rest of this paper.

Large deviation type estimates were introduced to study quasi-periodic Schr\"odinger operators in the late 1990s in a series of papers by Bourgain, Goldstein and Schlag, \cite{BG, GS1}. 
Their method has been well developed ever since and has shown to be sufficiently robust in the super-critical regime to deal with the following questions {{(not only restricted to the one dimensional quasi-periodic Schr\"odinger case)}}:
\begin{enumerate}
  \item Regularity of the  $L(E)$ and  $N(E)$ in energy $E$, (e.g. \cite{B2,GS1,BJ,K1,GS2}),
  \item Localization of the eigenfunctions, (e.g. \cite{BG,BGS,K1,GSV}),
  \item Eigenvalue separation and topological structure of the spectrum, (e.g. \cite{GS3,DGSV,GSV2}).
\end{enumerate}
In this paper, we will focus on Problem 1. For more details about Problems 2 and 3, we refer readers to \cite{GSsurvey,B1,JM17,DKbook} and references therein.


Proving regularity of $L(E)$ and $N(E)$ (in $E$) is considered difficult for any type of sequence of potentials, see \cite{CFKS}. 
Some weak regularity for general ergodic families was first proved in  cf~\cite{CS}.   
For quasi-periodic Schr\"odinger operators,  the first breakthrough was made by Goldstein and Schlag in \cite{GS1}.
They developed a robust scheme, by combining LDT with Avalanche Principle (AP), see Theorem \ref{thm:AVP}, to study the regularity problem.
They proved H\"{o}lder regularity of $L(E)$ and $N(E)$ for typical frequencies in $\T$, assuming analyticity of the potential and positive Lyapunov exponents. 
Some weaker H\"{o}lder regularity was also obtained in the same paper for $ \T^d$ with $d>1$. 
Bourgain and Jitomirskaya proved in \cite{BJ} that $L(\omega, E)$ is jointly-continuous in $(\omega, E)$ at any irrational $\omega\in \T$ for analytic potentials, 
this result was obtained by Bourgain for $\T^d$ with $d>1$ in \cite{BTd}.
More delicate estimates on sharp H\"older regularity for $\T$ were obtained by Goldstein and Schlag in \cite{GS2}.  
In a recent monograph by Duarte-Klein \cite{DKbook}, this scheme was extended systematically in depth and breadth, making it applicable to general cocycles, 
provided appropriate LDT estimates are available in the given setting.

For a general quasi-periodic analytic cocyle $(\omega, A)$, 
where $A$ is an analytic $m\times m$  matrix-valued function, 
regularity of $L(\omega, A)$ is formulated in terms of the analytic norm of $A$. 
Joint-continuity of $L(\omega, A)$ in $(\omega, A)$, without a modulus of continuity, was obtained in \cite{JMarxLE, AJS} at any irrational $\omega\in \T$. The approaches of \cite{JMarxLE, AJS} do not rely on LDT.
H\"older regularity of $L(\omega, A)$ was obtained by LDT in \cite{DKbook, DKJEMS} for Diophantine $\omega\in \T^d$, under the gapped Lyapunov exponent assumption (equivalent to positive (maximal) Lyapunov exponent when $m=2$).

In the sub-critical regime with analytic potential, regularity results were proved often by reducibility method, cf~\cite{A,AJ}.  
In the low-regularity potential regime, fewer results were obtained with more restrictions on the potential and the frequency, see for example \cite{K1, KlocG, A, WZ1, CCYZ,LWY}. 

In this paper we follow the scheme developed by Goldstein and Schlag \cite{GS1},
namely by combining LDT and AP to obtain the H\"older continuity of $L(E)$ and $N(E)$:
\begin{equation}\label{eq:holderGS01}
  |L(E)-L(E')|+|N(E)-N(E')|\le |E-E'|^\tau, \ \ |E-E'|\ll 1.
\end{equation}
One of their key estimate for the one-dimensional case is
\begin{equation}\label{eq:LDTGS01}
   {\rm mes}\big\{x\in\T:|u_n(\omega, E;x)-L_n(\omega, E)|> \kappa L(\omega, E)\big\}\le e^{-c(\omega,v,\kappa)L^2(\omega, E)n},
\end{equation}
under the positive Lyapunov exponent condition $L(\omega,E)>\gamma>0$, 
for $\omega$ satisfying the strong Diophantine condition, see \eqref{def:SDC}. 
However, due to the $L^2(\omega, E)$ term in the exponential estimate on the right hand side of \eqref{eq:LDTGS01}, 
the H\"older exponent $\tau$ in \eqref{eq:holderGS01} will tend to $0$ as the lower bound $\gamma$ approaches $0$.

In \cite{B1}, the LDT estimate (\ref{eq:LDTGS01}) was improved to be 
\begin{equation}\label{eq:LDTB1}
   {\rm mes}\big\{x\in\T:|u_n(\omega, E;x)-L_n(\omega, E)|> \kappa L(\omega, E)\big\}\le e^{-c(\omega,v,\kappa)L(\omega, E)n},
\end{equation}
in the small Lyapunov exponent regime, under the same assumption on $\omega$ \footnote{Note that we use the same symbol $c(\omega, v, \kappa)$ in both \eqref{eq:LDTGS01} and \eqref{eq:LDTB1}, but they are not the same constants.}. 
The improvement implies that the local H\"older exponent is independent of the lower bound $\gamma$.

As we mentioned above, both (\ref{eq:LDTGS01}) and (\ref{eq:LDTB1}) were established for $\omega$ satisfying a strong Diophantine condition (S.D.C.), which we define later. 
Going beyond S.D.C. is considered difficult for establishing LDT and H\"older continuity of the Lyapunov exponent in general.  
Our first result of this paper extends the LDT estimates to more frequencies in the best possible regime, see \eqref{ourcondition}.
Indeed, H\"older continuity fails for generic $\omega$, see \cite{AJ} \footnote{See the paragraph below Theorem 1.2 of \cite{AJ}: for $v=\lambda \cos$ with $\lambda\neq 0$, Lyapunov exponent is discontinuous at rational $\omega$'s, thus it is not H\"older for $\omega$'s that are well approximated by rationals.}.  Thus, 
the exponential decay (\ref{eq:LDTGS01}) or (\ref{eq:LDTB1}) cannot hold for all frequencies.

In both (\ref{eq:LDTGS01}) and (\ref{eq:LDTB1}), the dependence of $c(\omega,v,\kappa)$ on $v$ are not written down explicitly.
In our paper, we incorporate a refined Riesz-representation of subharmonic functions of \cite{GS2} into the proof of the LDT estimates.
This leads to an explicit dependence of $c$ on $v$.
It turns out that the constant depends on the potential $v$ in a ``$\sup-\sup$'' form, see \eqref{def:Cvrho}. 
If $v=\lambda f$, the ``$\sup-\sup$'' yields a magical cancellation of $\lambda$.
This leads to the second result of our paper, see Corollary \ref{cor:LDTlambda}.
Combining with AP, we obtain, for the first time, a $\lambda$-\emph{independent} H\"older exponent in the large coupling regime for general non-trivial analytic potentials, see Theorem \ref{thm:holder2}.
Such kind of result was previously only known for trigonometric polynomials.

In order to formulate our results, we introduce the following notations:
for any $x\in \R$, let $\|x\|_{\T}:=\inf_{n\in \Z} |x-n|$.
For any $\omega\in [0,1]
 \setminus \Q$, let $[a_1, a_2, a_3,...]$ be its continued fraction expansion.
Let $\{p_s/q_s\}_{s=1}^{\infty}$ be its continued fraction approximants, defined by $p_s/q_s=[a_1, a_2,..., a_s]$. 
It is well known that $\|q_s\omega\|_{\T}\le q_{s+1}^{-1}$. 
We say that $\omega$ satisfies a \emph{strong Diophantine condition (S.D.C.)} (or $\omega$ is strongly Diophantine)
\footnote{
We say that $\omega$ satisfies a Diophantine condition (D.C.) if
$\|n\omega\|_{\T}\ge\frac{c}{n^{a}}$ for all $n> 1$ and some $a>1,c>0$.
Note that for any $a>1$, a.e. $\omega$ satisfies a D.C. with some $c=c(\omega)>0$.
}
, if for some constants $a>1$, $c>0$ , the following holds for any $n\geq 1$,
\beq\label{def:SDC}
\|n\omega\|_{\T}\ge\frac{c}{n(1+\log n)^a}.
\eeq 
Note that for any $a>1$, a.e. $\omega$ satisfies S.D.C. for some $c=c(\omega)>0$.
It is also clear from the definition of S.D.C. that for strong Diophantine $\omega$,
\beq\label{eq:SDC}
q_{s+1}\leq c^{-1} q_s (\log{q_s})^a.
\eeq
Next we introduce an exponential growth exponent $\beta$ defined as follows:
\begin{equation}\label{def:beta}
  \beta(\omega):=\limsup_{s\rightarrow+\infty}\frac{\log q_{s+1}}{q_s}\in [0,\infty].
\end{equation}
It is then clear from \eqref{eq:SDC} that S.D.C.$\subsetneq$ $\{\omega:\beta(\omega)=0\}$. 
Those $\omega$ with $\beta(\omega)>0$ are usually called \emph{Liouville numbers}.

Since our potential $v(x)$ is a real analytic function, it has a bounded extension to a strip $|{\rm Im}z|<\rho$ with width denoted by $\rho>0$.
Let ${\cal N}_v=[-2-\|v\|_\infty,2+\|v\|_\infty]$ be the numerical range of the Schr\"odinger operator $H$. 
It is well known that $\sigma(H)\subset {\cal N}_v$ and $L(E)$ is a $C^{\infty}$ function outside of the spectrum. 
Hence we will only consider $E\in {\cal N}_v$ throughout the paper.

\begin{thm}\label{thm:LDTmain}
Let $\omega\in \R\setminus \Q$. 
There exist constants $c({v, \rho}), \tilde{c}({v, \rho})\in(0,1)$ such that,
if 
\beq\label{ourcondition}
0\le \beta(\omega)<c(v,\rho) \inf_{E\in [a,b]} L(\omega, E), 
\eeq 
then there is
$N=N(\omega, \inf_{E\in [a, b]} L(\omega, E), v, \rho) \in \N$ such that for any $n\geq N$ the following large deviation estimates hold uniformly in $E\in [a, b]$,
\begin{description}
  \item[(a)] If $0<L(\omega,E)<1$, then
  \begin{equation}\label{eq:LDTsmall}
    {\rm mes}\left\lbrace x\in\T:\mid u_n(\omega, E; x)-L_n(\omega, E)\mid>\frac{1}{20}L(\omega, E) \right\rbrace \leq e^{-\tilde{c}(v,\rho) L(\omega, E)\,n}.
  \end{equation}
  \item[(b)]  If $L(\omega,E)\ge1$, then
  \begin{equation}\label{eq:LDTlarge}
    {\rm mes}\left\lbrace x\in\T:\mid u_n(\omega, E; x)-L_n(\omega, E)\mid>\frac{1}{20}L(\omega, E) \right\rbrace \leq e^{-\tilde{c}(v,\rho) L^2(\omega, E)\,n}.
  \end{equation}
\end{description}
\end{thm}

\begin{rmk}
The parameter $1/20$ in Theorem \ref{thm:LDTmain} can be replaced by any $0<\kappa<1$. 
The new constants $c_\kappa({v,\rho}), \tilde{c}_\kappa({v, \rho})$ only differ from $c({v,\rho}), \tilde{c}({v, \rho})$ by a constant multiple of $\kappa^{2}$. 
However, in order to apply AP to obtain H\"older continuity, $\kappa$  can be taken to be at most $1/9$ due to technical reasons
(see (\ref{eq:AvpBj})). 
We do not intend to improve the H\"older exponents in the paper by getting the best possible $\kappa$, thus we take $\kappa=1/20$ for simplicity. See more discussions about the sharp H\"older exponents after Theorem \ref{thm:holder1}.
\end{rmk}

\begin{cor}\label{cor:LDTlambda}
Let $\omega\in \R\setminus \Q$.
Assume that $v(x)$ in (\ref{def:Schop}) is given by $v(x)=\lambda f(x)$, where $\lambda$ is a positive constant. 
There exist constants  $0<b=b(f,\rho)<1,B=B(f,\rho)>1$ and $\tilde{\lambda}=\tilde{\lambda}(f,\rho)>0$ with the following properties:
for any irrational $\omega$ with $0\le\beta(\omega)<\infty$, suppose
$$\lambda>\max(\tilde \lambda,e^{B\beta(\omega)}),$$ 
 then there is $N(\omega,\lambda, f, \rho)\in \N$ such that for any $n\geq N(\omega,\lambda, f, \rho)$, the following holds
\begin{align}\label{eq:LDTlambda}
{\rm mes}\left\lbrace x\in\T:\mid u_n(\omega, E; x)-L_n(\omega, E)\mid>\frac{1}{19}\log{\lambda} \right\rbrace
\leq e^{-n\,b\log{\lambda}}.
\end{align}
\end{cor}

\begin{rmk}
The above exponential decay of the measure estimate  w.r.t. $\log \lambda$ for large coupling $\lambda$ is known for the first time even for $\beta(\omega)=0$ or S.D.C. $\omega$ to the authors' knowledge.
\end{rmk}

As mentioned previously in (\ref{eq:holderGS01}), a direct consequence of the above large deviation estimates is the H\"older regularity of the Lyapunov exponents. With the refined parameters in the LDT estimates (\ref{eq:LDTsmall})-(\ref{eq:LDTlambda}), we have the following H\"older continuity of  the Lyapunov exponents.

\begin{thm}\label{thm:holder1}
Let $c=c({v, \rho}), \tilde c=\tilde{c}({v, \rho})$ be the constants in Theorem \ref{thm:LDTmain}.
There exists a constant $\tau>0$ depending explicitly (and only) on $\tilde{c}({v, \rho})$ that satisfies the following property:
if $(\omega_0, E_0) \in (\R\setminus \Q)\times{\cal N}_v$ is a point with $L(\omega_0,E_0)=\gamma>0$, 
and $U\times I$ is a neighborhood of $(\omega_0, E_0)$ such that $L(\omega, E) \in [\frac{18}{19}\gamma, \frac{20}{19}\gamma]$,
then for any $\omega\in U$ with 
\begin{align*}
0\leq \beta(\omega)<\frac{1}{2}c\,\gamma,
\end{align*} 
there is $\eta=\eta(\omega,I,\gamma,v)$
such that the following holds for any $E, E'\in I$ and $|E-E'|<\eta$,
\begin{equation}\label{eq:holder1}
|L(\omega,E)-L(\omega,E')|\le |E-E'|^{\tau}.
\end{equation}
\end{thm}

\begin{rmk}\label{rmk:QIexists}
By \cite{BJ}, $L(\omega, E)$ is jointly-continuous in $(\omega, E)$ at $(\omega_0, E_0)$.
Hence the neighborhood $U\times I$ always exists.
\end{rmk}

\begin{rmk}
Theorem \ref{thm:holder1} shows that the exponent $\tau$ is independent of the lower bound of the Lyapunov exponent $\gamma$.
This generalizes the result in \cite{B1} for general analytic potentials from $\omega$ satisfying S.D.C. to $0\leq \beta(\omega)\lesssim \gamma$. 
\end{rmk}

\begin{rmk}
For trigonometric polynomial potentials, there are results on sharp H\"older exponents that only depend on the degree of the polynomial: 
$\frac{1}{2}$-H\"older if $v=\lambda \cos$, $\lambda\neq 0,1$, \cite{B2, AJ}; 
and $(\frac{1}{2k}-\epsilon)$-H\"older if $v$ is a small $C^\infty$ perturbation of a trigonometric polynomial of degree $k$ \cite{GS2}. 
Our current approach does not lead to such kind of sharp exponent for general analytic potentials, even for S.D.C. $\omega$. 
\end{rmk}

\begin{rmk}
If $v$ is of the form $\lambda f$, with a general analytic $f$, in the small coupling regime $\lambda<\lambda_0(f)$, $\frac{1}{2}$-H\"older exponents were obtained in \cite{AJ} using a reducibility method. However there is no such kind of result for the large coupling regime
	\footnote{{For general analytic potential $v=\lambda f$, if one applies the LDT (\ref{eq:LDTGS01}) in \cite{GS1} and check all the constants explicitly, the H\"older exponent behaves like $O({(\log \lambda)^{-1}})$ for large $\lambda$ even for S.D.C. $\omega$, see more explanation in \cite{YZ}. }}. Our Theorem \ref{thm:holder2} is the first one in this regime, by giving a $\lambda$-independent H\"older exponent for general analytic $f$. 
\end{rmk}

If $v=\lambda f$, we have the following:
\begin{thm}\label{thm:holder2}
Under the same condition of Corollary \ref{cor:LDTlambda}, let $\tilde{\lambda}(f,\rho), b(f,\rho), B(f,\rho)$ be the constants given there.
There exists a constant $\tilde \tau>0$ depending explicitly (and only) on $b$ (hence independent of $\lambda$) such that
for any irrational $\omega$ with $0\leq \beta(\omega)<\infty$, if $\lambda>\max(\tilde{\lambda},e^{B\beta(\omega)})$, then there exists $\tilde{\eta}=\tilde{\eta}({\omega,\lambda,f,\rho})>0$, such that for any $E,E'\in{\cal N}_{\lambda f}$ and  $|E-E'|\le \tilde\eta$, we have
 \begin{equation}\label{eq:holder2}
   |L(E)-L(E')|\le |E-E'|^{\tilde\tau}. 
 \end{equation}
\end{thm}

The rest of the paper is organized as follows: in section \ref{sec:lemma}, we state all the important technical  lemmas. In section \ref{sec:proofofLDT}, we prove the three large deviation estimates using the lemmas in Section \ref{sec:lemma}. Our H\"older continuity follows directly from LDT and a standard argument combined with the Avalanche Principle. For the sake of completeness, we sketch the proof in section \ref{sec:proofofHolder}. Many details of this part are included in the Appendix for reader's convenience.

\noindent {\bf Acknowledgment.} 
R. H. is grateful to Wilhelm Schlag for introducing the Riesz-representation and LDT to him.
Both authors would like to thank Wilhelm Schlag for useful discussions.
R. H. would like to thank the Institute for Advanced Study, Princeton, for its hospitality during the 2017-18 academic year.
This material is based upon work supported by the National Science Foundation under Grant No. DMS-1638352.
Research of S. Z. was supported in part by NSF grant DMS-1600065 and DMS-1758326.

\section{Useful lemmas}\label{sec:lemma}
Let ${\cal N}_v=[-2-\|v\|_\infty,2+\|v\|_\infty]$, as we mentioned before, we will only consider $E\in{\cal N}_v$ throughout the paper.
Recall that $u_n(\omega,E;x)$ is defined as in \eqref{def:unLn}.

This section contains lemmas that will be used in the proofs of Theorems \ref{thm:LDTmain} and \ref{cor:LDTlambda}.
The proofs of these lemmas will be included in Sec. \ref{sec:proofoflemma}.

Let
\beq\label{def:Cv}
\Lambda_v:=\log{(3+2\|v\|_{L^{\infty}(\T)})}.
\eeq
Simple computations yield that 
\begin{align}\label{trivialbdd}
\sup_{E\in {\cal N}_v}\lv u_n(\omega, E; \cdot)\rv_{L^\infty(\T)}\leq \Lambda_v,
\end{align}
holds uniformly in $\omega\in \T$ and $1\leq  n\in \N$.

Since in our model, $v$ is assumed to have bounded analytic extension to $\T_\rho:=\{z:|{\rm Im}z|<\rho\}$, $u_n$ has subharmonic extension on $\T_\rho$ with a uniform upper bound 
\begin{align*}\label{}
\sup_{E\in {\cal N}_v}\sup_{n\in\N}\lv u_n(\omega, E; \cdot)\rv_{L^\infty(\T_\rho)}\leq \log{(3+2\|v\|_{L^{\infty}(\T_\rho)})}<\infty. 
\end{align*}

\subsection{Estimates of the Fourier coefficients $\hat{u}_n(\omega, E; k)$}
The function $u_n(\omega,E;x)$ is  1-periodic  on $\R$ and we denote its Fourier coefficients by
\begin{equation}\label{def:hatunk}
\hat{u}_n(\omega, E; k)=\int_\T u_n(\omega,E;x)e^{-2\pi{i}kx}{\rm d}x.
\end{equation}
The following estimate of the Fourier coefficient is well-known and crucial to establishing our LDT, see e.g. Bourgain's monograph \cite[Corollary 4.7]{B1}.
For a version of this estimate written precisely in the ``$\sup-\sup$'' form below, see \cite[Lemma 2.8]{DK1}.
To obtain this ``$\sup-\sup$'' estimate, one needs to invoke a refined Riesz-representation theorem \cite[Lemma 2.2]{GS2}.
See details in Sec. \ref{sec:proofukold}.
\be{lm}\label{lm:ukold}
There is a constant $\alpha(\rho)>0$ depending on $\rho$ only, 
such that for any $k\neq 0$,
\begin{align}\label{eq:ukold}
\lvert \hat{u}_n(\omega, E; k)\rvert \leq \frac{\alpha(\rho)}{|k|}\Big(\sup_{|{\rm Im}z|<\rho }u_n(\omega, E; z)-\sup_{|{\rm Im}z|<\rho/2 }u_n(\omega, E; z)\Big).
\end{align}
\e{lm}

\be{cor}\label{cor:ukolduniform}
Let 
\begin{align}\label{def:Cvrho}
C(v,\rho):=\alpha(\rho) \sup_{E\in{\cal N}_v}\Big(\sup_{|{\rm Im}z|<\rho }u_n(\omega, E; z)-\sup_{|{\rm Im}z|<\rho/2 }u_n(\omega, E; z)\Big)<\infty.
\end{align}
We then have that for any $k\neq 0$ and $E\in{\cal N}_v$,
\begin{align}\label{eq:ukolduniforminE}
\lvert \hat{u}_n(\omega, E; k)\rvert \leq \frac{C(v,\rho)}{|k|}.
\end{align}
\e{cor}

When $v$ is given as $\lambda f$, we can bound the above constant $C(\lambda f,\rho)$ by a constant independent of $\lambda$. 
This turns out to be crucial to our proof of Corollary \ref{cor:LDTlambda}.
\begin{lm}\label{lm:uniformsup-sup}
Let $C(v,\rho)$ be the constant defined  in \eqref{def:Cvrho}. Suppose that $v=\lambda f$. Then there is $C_0(f,\rho)>0$, independent of $\lambda$, such that for any $\lambda>0$,
\begin{equation}\label{eq:uniformsup-sup}
C({\lambda f, \rho})\le C_0(f, \rho).
\end{equation}
\end{lm}

Besides the Fourier decay estimate in Lemma \ref{lm:ukold}, we also prove a new estimate as follows. 
This estimate improves that of Lemma \ref{lm:ukold} for small $|k|$ when $n$ is large. 
It will play a crucial role in our proof of part (a) of Theorem \ref{thm:LDTmain}.
\be{lm}\label{lm:uknew}
Let $\Lambda_v$ be the constant defined in \eqref{def:Cv}.
We have the following bounds of the Fourier coefficients, for any $k\neq 0$,
\begin{align*}
\lvert \hat{u}_n(\omega, E; k)\rvert \leq \frac{\Lambda_v}{2n \|k\omega\|_{\T}}.
\end{align*}
\e{lm}

\subsection{$\|u_n(\omega, E; \cdot)\|_{L^\infty(\T)}$ under small Lyapunov exponent condition}

We present an upper bound of $\|u_n(\omega, E; x)\|$, see Lemma \ref{lm:unupper} below.
This can be viewed as a generalization of \cite[Lemma 8.18]{B1}, where a similar bound was proved for Diophantine $\omega$.
Compared to a trivial bound $\|u_n(\omega, E;x\|\leq \Lambda_v$, the new bound is much more effective when the Lyapunov exponent is small.

Compared to \cite[Lemma 8.18]{B1}, our improvement lies in the fact that we can relax the Diophantine condition on $\omega$. 
Indeed we give explicit dependence of the upper bound on the continued fraction approximants of $\omega$, through the $\log{q_{s+1}}/q_s$ term.
This improvement enables us to cover Liouville frequencies.

For $R\in \N$, let $u_n^{(R)}$ be the average of $u_n$ along a trajectory with length $\sim R$, defined as:
\beq\label{def:uR}
u_n^{(R)}(\omega, E; x):=\sum_{|j|<R}\frac{R-|j|}{R^2} u_n(\omega, E; x+j\omega).
\eeq

\be{lm}\label{lm:uRupper}
Let $C(v,\rho)$, $C_3$ be the constants in \eqref{def:Cvrho}, \eqref{eq:udeltak3}.
Assuming that $0<L(\omega, E)<1$, we have the following upper bound of $u_n^{(R)}(\omega, E; x)$,
\begin{align}\label{eq:uRupper1}
\|u_n^{(R)}(\omega, E; \cdot)\|_{L^\infty(\T)}\leq 
L_n(\omega, E)
+(2+8 C(v, \rho)+4\pi C_3 C(v,\rho)) L(\omega, E)
+120 C(v, \rho) \frac{\log{q_{s+1}}}{q_s},
\end{align}
which holds for
$$n\geq 2\Lambda_v L(\omega, E)^{-2} \sup_{1\leq |k|\leq L(\omega, E)^{-1}}\frac{1}{\|k\omega\|_{\T}},\ \ \
\text{and}\ \ \ R\geq 144 L(\omega, E)^{-5}.$$
\e{lm}

Lemma \ref{lm:uRupper} leads to the following
\be{lm}\label{lm:unupper}
Let $C(v,\rho)$, $C_3$ be the constants in \eqref{def:Cvrho}, \eqref{eq:udeltak3}.
Assuming that $0<L(\omega, E)<1$, we have the following upper bound of $u_n(\omega, E; x)$,
\begin{align}\label{eq:unupper1}
\|u_n(\omega, E; \cdot)\|_{L^\infty(\T)}\leq L_n(\omega, E)+C_1 L(\omega, E)
+120 C(v, \rho) \frac{\log{q_{s+1}}}{q_s},
\end{align}
which holds for $n\geq N_0(\omega, L(\omega, E), v, \rho)$, where $C_1$ explicitly depends on $C({v, \rho}), {\Lambda_v}$ as
\begin{equation}\label{def:C2}
C_1:=2+\Lambda_v+8 C(v, \rho) +4\pi C_3 C(v,\rho)
\end{equation}
and
\beq\label{def:N0}
\begin{aligned}
N_0(\omega, L(\omega, E), v, \rho):=L(\omega, E)^{-2}  \max\left( 2\Lambda_v  \sup_{1\leq |k|\leq L(\omega, E)^{-1}}\frac{1}{\|k\omega\|_{\T}},\ \ 
49 L(\omega, E)^{-4} \right).
\end{aligned}
\eeq
\e{lm}

\subsection{Two estimates of $\|u_n(\omega, E; \cdot)-u^{(R)}_n(\omega, E; \cdot)\|_{L^\infty(\T)}$}

The following lemmas give upper bounds of $\|u_n-u^{(R)}_n\|_{L^\infty(\T)}$ under different conditions.

\be{lm}\label{lm:un-uRn0}
Let $\Lambda_v$ be the constant defined in \eqref{def:Cv}.
For any $n,R,\omega$, we have
$$\lv u_n(\omega, E; \cdot)-u_n^{(R)}(\omega, E; \cdot)\rv_{L^\infty(\T)}\le 2\Lambda_v \frac{R}{n}.$$
\e{lm}

Recall the following uniform convergence in \cite{BJ}.
\begin{lm}\cite[Corollary 3]{BJ}\label{thm:bj02uniform}
Suppose $v$ is analytic. Then
\begin{equation}\label{eq:bj02uniform}
  \limsup_{n\to\infty}u_n(\omega,E;x) \le L(\omega,E)
\end{equation}
uniformly in $x$ and $E$ in a compact set.
\end{lm}
A direct consequence is
\be{lm}\label{lm:uniformtildeN0}
Suppose $L(\omega,E)>0$ for all $E\in[a,b]$. There exists $\widetilde{N}_0(\omega, [a, b], v)$ such that for any $n>\widetilde{N}_0(\omega, [a, b], v)$, any $x\in \T$ and $E\in [a,b]$, we have
\begin{equation}\label{eq:uniformUN}
  u_n(\omega,E;x)\le \left(1+\frac{1}{20}\right)L(\omega, E)
\end{equation}
and
\begin{equation}\label{eq:uniformLN}
  L_n(\omega,E)\le \left(1+\frac{1}{20}\right)L(\omega, E).
\end{equation}
\e{lm}

A more delicate upper bound of the difference $u_n-u_n^{(R)}$, when $L(\omega, E)$ is small, is given as follows. 
This upper bound will be the key to Theorem \ref{thm:LDTmain}, part (a).
Let $N_0$ be as in \eqref{def:N0} and $\widetilde N_0$ be as in Lemma \ref{lm:uniformtildeN0}.
Define
\begin{align}\label{def:N1}
N_1(\omega, [a, b], L(\omega, E), v, \rho):=\max(N_0(\omega, L(\omega, E), v, \rho),\ \widetilde{N}_0(\omega, [a, b], v)+1).
\end{align}

Using Lemma \ref{lm:unupper} and Lemma \ref{lm:uniformtildeN0}, we obtain the following:
\be{lm}\label{lm:un-uRn}
Let $C_1,N_1$ be the constants in \eqref{def:C2},\eqref{def:N1} and $C(v,\rho)$, $\Lambda_v$ be the constants in \eqref{def:Cvrho}, \eqref{def:Cv} respectively.
Suppose $0<L(\omega, E)<1$. For $R=\lfloor \left(400 \left(C_1+2 \right) \right)^{-1} n\rfloor+1$, we have that 
\begin{align}\label{eq:unupper1}
\lv u_n(\omega, E; \cdot)-u^{(R)}_n(\omega, E; \cdot)\rv_{L^\infty(\T)}\leq \frac{1}{100}L(\omega, E)\nonumber
+\frac{1}{5}C({v, \rho})\frac{\log{q_{s+1}}}{q_s},
\end{align}
holds for $n\geq  N_2(\omega, [a, b], L(\omega, E), v, \rho)$, where 
\beq\label{def:N2}
N_2(\omega, [a, b], L(\omega, E), v, \rho):=
\max{(150 \Lambda_v N_1 L(\omega, E)^{-1},\ 400 (C_1+2) N_1+1 )}.
\eeq
\e{lm}
\be{rmk}\label{rmk:N2de}
We point out that $N_1(\omega, [a, b], L(\omega, E), v, \rho)$ is a decreasing function in the third parameter $L(\omega, E)$, and so is $N_2(\omega, [a, b], L(\omega, E), v, \rho)$. This is clear from the definitions \eqref{def:N0}, \eqref{def:N1} and \eqref{def:N2}.
\e{rmk}

\section{Large deviation estimates.}\label{sec:proofofLDT}
For simplicity, from this point on, when there is no ambiguity, we will sometimes write $u_n(x)=u_n(\omega, E; x)$, $L_n=L_n(\omega, E)$ and $L=L(\omega, E)$. 

\subsection{Preparation}
Let $\hat u_n(k)$ and $u_n^{(R)}(x)$ be defined as in (\ref{def:hatunk}),(\ref{def:uR}). Let
\begin{equation}\label{def:FRk}
F_R(k):=\sum_{|j|<R}\frac{R-|j|}{R^2}e^{2\pi { i}kj\omega}.
\end{equation}
Let us recall the following estimates of $F_R(k)$ in \cite{BJ,B1,YZ}, whose proofs are included in the Appendix \ref{sec:proofofFRk}.
\begin{align}
0\leq &F_R(k)\le \min \left(1,\ \frac{2}{1+R^2\|k\omega\|_{\T}^2}\right),\label{eq:FRkleq}\\
\sum_{1\leq |k|<q/4} & \frac{1}{1+R^2 \|k\omega\|_{\T}^2}\leq 2\pi \frac{q}{R}, \label{eq:lowqsR}\\
\sum_{|k|\in [\ell q/4,\ (\ell+1) q/4)} & \frac{1}{1+R^2 \|k\omega\|_{\T}^2}\leq 2+4\pi \frac{q}{R},\ \ \forall \ell\in \N, \label{eq:low1+qsR}
\end{align}
in which $p/q$ is any continued fraction approximant of $\omega$.

Direct computation shows that
\begin{equation}\label{eq:unRFR}
u_n^{(R)}(x)=L_n+\sum_{k\in\Z,k\neq0}\hat{u}_n(k)F_R(k)e^{2\pi { i}kx}
\end{equation}

Let ${p_s}/{q_s}$, ${p_{s+1}}/{q_{s+1}}$ be any two consecutive continued fraction approximants of $\omega$.
For $0<\delta\le 1$, let us consider
\beq\label{eq:U0}
\begin{aligned}
u_n(x)-L_n
=&u_n(x)-u_n^{(R)}(x)+u_n^{(R)}(x)-L_n\\
=&u_n(x)-u_n^{(R)}(x)  \ \ (=:\mathcal{U}_1(x))\\
&\ \ +\sum_{1\leq |k|< \delta^{-1}} \hat{u}_n(k) F_R(k) e^{2\pi i kx} \ \ (=:\mathcal{U}_2(x))\\
&\ \ +\sum_{\delta^{-1}\leq |k|< q_s/4} \hat{u}_n(k) F_R(k) e^{2\pi i kx} \ \ (=:\mathcal{U}_3(x))\\
&\ \ +\sum_{q_s/4\leq |k|<q_{s+1}/4} \hat{u}_n(k) F_R(k) e^{2\pi i kx} \ \ (=:\mathcal{U}_4(x))\\
&\ \ +\sum_{q_{s+1}/4\leq |k|<K} \hat{u}_n(k) F_R(k) e^{2\pi i kx} \ \ (=:\mathcal{U}_5(x))\\
&\ \ + \sum_{|k|\geq K} \hat{u}_n(k) F_R(k) e^{2\pi i kx} \ \ (=:\mathcal{U}_6(x)).
\end{aligned}
\eeq

By Lemma \ref{lm:uknew}, we have some refined estimates of $\mathcal{U}_2(x)$ and $\mathcal{U}_3(x)$:
\begin{prop}\label{u2u3}
Let $\Lambda_v,C({v,\rho})$ be given as in (\ref{def:Cv}), (\ref{def:Cvrho}). For any $n\ge1$ and $R\in[q_s,q_{s+1})$, we have
\begin{equation}\label{u2}
  \|\mathcal{U}_2(\cdot)\|_{L^\infty(\T)}\le
  \frac{\Lambda_v}{\delta n }\cdot \sup_{1\leq k\leq \delta^{-1}} \frac{1}{\|k\omega\|_{\T}}
\end{equation}
and
\begin{equation}\label{u3}
  \|\mathcal{U}_3(\cdot)\|_{L^\infty(\T)}
\leq 4 \pi \delta C({v, \rho})
\end{equation}
\end{prop}
\noindent \textbf{Proof:}
By Lemma \ref{lm:uknew} {and \eqref{eq:FRkleq}}, we have
\beq\label{eq:U2delta}
\begin{aligned}
\|\mathcal{U}_2(\cdot)\|_{L^\infty(\T)}
\leq \sum_{1\leq |k|< \delta^{-1}} \lvert  \hat{u}_n(k)\rvert
\leq \frac{\Lambda_v}{2n}  \sum_{1\leq |k|< \delta^{-1}}\frac{1}{\|k\omega\|_{\T}}
\leq \frac{\Lambda_v}{\delta n} \sup_{1\leq k\leq \delta^{-1}} \frac{1}{\|k\omega\|_{\T}}.
\end{aligned}
\eeq

By Lemma \ref{lm:ukold}, \eqref{eq:FRkleq}, \eqref{eq:lowqsR} and $q_s\le R$, we obtain
\beq\label{eq:U3delta}
\begin{aligned}
\|\mathcal{U}_3(\cdot)\|_{L^\infty(\T)}
\leq &2\sum_{\delta^{-1}\leq |k|< q_s/4} \lvert\hat{u}_n(k)\rvert \frac{1}{1+R^2\|k\omega\|_{\T}^2}
\leq 2  \sum_{\delta^{-1}\leq |k|< q_s/4}\frac{C({v, \rho}) }{\delta^{-1}}\frac{1}{1+R^2\|k\omega\|_{\T}^2}\\
\leq &2 C({v, \rho}) \cdot\delta \cdot\sum_{1\leq |k|< q_s/4}\frac{1}{1+R^2\|k\omega\|_{\T}^2}
\leq 4\pi C({v, \rho})\cdot \delta\cdot \frac{q_s}{R}
\leq 4\pi \delta C({v, \rho}),
\end{aligned}
\eeq
as desired.
\qed

We have some general estimates for $\mathcal{U}_4(x)+\mathcal{U}_5(x)$ and $\mathcal{U}_6(x)$.
\begin{prop}\label{prop:u4u5u6}
Let $C({v,\rho})$ be given as in (\ref{def:Cvrho}). For any $n\ge1$, and $q_s\leq R<q_{s+1}\leq K$, we have

\beq\label{eq:u4u5}
\begin{aligned}
 \|\mathcal{U}_4(\cdot)+\mathcal{U}_5(\cdot)\|_{L^\infty(\T)}
\leq & 120 C({v,\rho}) \left(\frac{\log{q_{s+1}}}{q_s}+\frac{\log{K}}{R}\right),  
\end{aligned}
\eeq
and
\begin{equation}\label{eq:u6}
  \|\mathcal{U}_6(\cdot)\|_{L^2(\T)}^2
\leq C^2({v,\rho})\frac{2}{K}.
\end{equation}
\end{prop}
This part has been proved in \cite{YZ}, but we sketch the proof below for reader's convenience. 

\noindent \textbf{Proof:}
By Lemma \ref{lm:ukold}, \eqref{eq:FRkleq}, \eqref{eq:low1+qsR} and the choice of $R\in[q_s,q_{s+1})$, we have
\beq\label{U4}
\begin{aligned}
\|\mathcal{U}_4(\cdot)\|_{L^\infty(\T)}
\leq &2\sum_{q_s/4\leq |k|<q_{s+1}/4}\lvert \hat{u}_n(k) \rvert \frac{1}{1+R^2\|k\omega\|_{\T}^2}\\
\leq &2\sum_{\ell=1}^{\lfloor q_{s+1}/q_s \rfloor+1} \sum_{|k|\in [\ell q_s/4, (\ell+1)q_s/4)}\lvert \hat{u}_n(k) \rvert \frac{1}{1+R^2\|k\omega\|_{\T}^2}\\
\leq &8C({v,\rho})\sum_{\ell=1}^{\lfloor q_{s+1}/q_s \rfloor+1} \sum_{|k|\in [\ell q_s/4, (\ell+1)q_s/4)} \frac{1}{\ell q_s}\cdot \frac{1}{1+R^2\|k\omega\|_{\T}^2}\\
\leq &8C({v,\rho})\sum_{\ell=1}^{\lfloor q_{s+1}/q_s \rfloor+1} \frac{1}{\ell q_s} \left(2+4\pi \frac{q_s}{R}\right)\\
\leq &16C({v,\rho})\left(1+2\pi \right) \frac{\log{q_{s+1}}}{q_s}.
\end{aligned}
\eeq


In view of $\mathcal{U}_5$, we have by Lemma \ref{lm:ukold} and \eqref{eq:low1+qsR} that
\beq\label{U5}
\begin{aligned}
\|\mathcal{U}_5(\cdot)\|_{L^\infty(\T)}
\leq &2\sum_{q_{s+1}/4\leq |k|\leq K} \lvert \hat{u}_n(k) \rvert \frac{1}{1+R^2\|k\omega\|_{\T}^2}\\
\leq &8C({v,\rho})\sum_{\ell=1}^{\lfloor 4K/q_{s+1}\rfloor+1} \sum_{|k|\in [\ell q_{s+1}/4, (\ell+1)q_{s+1}/4)} \frac{1}{\ell q_{s+1}}\cdot \frac{1}{1+R^2\|k\omega\|_{\T}^2}\\
\leq &16C({v,\rho})\left(1+2\pi\right)\frac{\log{K}}{R}.
\end{aligned}
\eeq

Combining \eqref{U4} with \eqref{U5}, and use that $16(1+2\pi)<120$, we prove \eqref{eq:u4u5}.

For $\mathcal{U}_6$, we have that by Lemma \ref{lm:ukold},
\beq\label{U6}
\begin{aligned}
\|\mathcal{U}_6(\cdot)\|_{L^2(\T)}^2 \leq \sum_{|k|> K} |\hat{u}_n(k)|^2
\leq C^2({v, \rho})\sum_{|k|> K} \frac{1}{k^2}
\leq &C^2({v, \rho}) \frac{2}{K},
\end{aligned}
\eeq  
as claimed.
\qed

\subsection{Proof of Theorem \ref{thm:LDTmain}}
Let 
\beq\label{def:underlineL}
\underline{L}(\omega, [a,b])=\inf_{E\in [a, b]} L(\omega, E),\ \ \text{and}\ \ \tilde{\underline{L}}(\omega, [a,b])=\min (\underline{L}(\omega, [a,b]), 1).
\eeq
For simplicity, we will sometimes omit the dependence on $\omega$ and $[a,b]$ and write $\underline{L}$ and $\tilde{\underline{L}}$ instead.

Recall our notations: $N_2$ as in \eqref{def:N2}, and $\Lambda_v, C(v,\rho), C_1$ 
		as in \eqref{def:Cv}, (\ref{def:Cvrho}) and (\ref{def:C2}).

We choose $c$ and $\tilde{c}$ in the statement of the theorem as follows:
\beq\label{def:ctildecThm(a)}
c(v,\rho)=\big(36000 C({v, \rho})\big)^{-1},\ \ \ 
\tilde{c}(v,\rho)=\big(2\times 10^7(C_1+2)C({v, \rho})\big)^{-1}.
\eeq

By our condition:
\beq\nn
\beta(\omega)=\limsup_{k\rightarrow\infty} \frac{\log{q_{k+1}}}{q_k}\leq c(v,\rho) \underline{L}(\omega).
\eeq
Hence there exists $s_0=s_0(\omega, [a,b], v,\rho)$ such that for any $k\geq s_0$,
\beq\label{def:s0}
\frac{\log{q_{k+1}}}{q_k}\leq 2 c(v,\rho) \underline{L}(\omega, [a,b]).
\eeq

Let $n\geq N$, with $N$ defined as follows:
\beq\label{def:Ncase1}
\begin{aligned}
N(\omega, \underline{L}, v, \rho):=
\max 
\begin{cases}
&(i).\ 400(C_1+2)q_{s_0},\\
\\
&(ii).\ N_2(\omega, [a,b], \underline{L}, v, \rho),\\ 
\\
&(iii).\ 1.6\times 10^5 \pi \Lambda_v C(v,\rho) \tilde{\underline{L}}^{-2} \sup_{1\leq k\leq 800\pi C(v, \rho) \tilde{\underline{L}}^{-1}} \frac{1}{\|k\omega\|_{\T}},\\
\\
&(iv).\ 2\times 10^7(C_1+2)C({v, \rho}) \tilde{\underline{L}}^{-1} \log\left(2\times10^4C^2({v, \rho}) \tilde{\underline{L}}^{-2}+e\right).
\end{cases}
\end{aligned}
\eeq
This gives four lower bounds of $n$.

\begin{rmk}\label{rmk:Nde}
By Remark \ref{rmk:N2de}, $N_2$ is decreasing in $\underline{L}$. It is also clear that both (iii) and (iv) are decreasing in $\underline{L}$. 
Hence $N$ is non-increasing in $\underline{L}$.
\end{rmk}

\subsubsection{Parameters for part (a)}
In this case, $L<1$, hence 
\beq\label{eq:L0=L}
\tilde{\underline{L}}=\underline{L}.
\eeq

In our decomposition of $u_n(x)-L_n$ in \eqref{eq:U0},
we choose the following parameters:
\beq\label{def:constantsThm1a}
\begin{aligned}
\delta &=\frac{\underline{L}}{800\pi C({v, \rho})},\ \ \ \ \ 
R=\left[\frac{n}{400(C_1+2)}\right]+1,\\
K&=\left[\exp{\left(\frac{RL}{1.2 \times 10^4 C({v, \rho})}\right)}\right],\ \ \ \ \ 
s=\max \left\lbrace s\in \N:\ q_s\leq R\right\rbrace.
\end{aligned}
\eeq
It is clear from the choice of $s$ that $q_s\leq R<q_{s+1}$.
Let us also note that with $\delta$ defined above, the lower bound (iii) in \eqref{def:Ncase1} becomes
\beq\label{eq:part(a)N(c)}
\frac{200\Lambda_v}{\delta \underline{L}} \sup_{1\leq k\leq \delta^{-1}} \frac{1}{\|k\omega\|_{\T}}.
\eeq

Indeed, by (i) of \eqref{def:Ncase1}, we have
\beq\nn
R>\left(400(C_1+2)\right)^{-1}n\geq q_{s_0}.
\eeq
By our definition of $s$, see \eqref{def:constantsThm1a}, we clearly have $s\geq s_0$.
This, by \eqref{def:s0}, implies
\beq\label{eq:betas<2cL}
\frac{\log{q_{s+1}}}{q_s}\leq 2c(v,\rho)  \underline{L}.
\eeq
An upper bound of $q_{s+1}$ could be derived from \eqref{eq:betas<2cL}. Indeed,
\beq\label{eq:qs+1<K1}
\begin{aligned}
q_{s+1}
\leq &\exp{\left(2c(v,\rho) \underline{L} q_s\right)}
\leq \exp{\left(2c(v,\rho) \underline{L} R\right)} 
\leq \exp{\left(\frac{L R}{1.8 \times 10^4 C(v,\rho)}\right)}.
\end{aligned}
\eeq
By (iv) of \eqref{def:Ncase1},
\beq\nn
\begin{aligned}
n
\geq &2\times 10^7 (C_1+2) C({v, \rho}) \underline{L}^{-1},
\end{aligned}
\eeq
hence, we have
\beq\nn
\begin{aligned}
\exp{\left(\frac{L R}{1.8\times 10^4 C(v,\rho)}\right)}
\geq \exp{\left(\frac{ \underline{L} n}{7.2\times 10^6 (C_1+2) C(v,\rho)}\right)}
\geq \exp{\left(\frac{2\times 10^7}{7.2 \times 10^6 }\right)}>16.
\end{aligned}
\eeq
Using the fact that 
$
x<x^{\frac{3}{2}}-1$, for  $x>3$,
we have
\beq\label{eq:qs+1<K2}
\begin{aligned}
\exp{\left(\frac{L R}{1.8 \times 10^4 C(v,\rho)}\right)}<\exp{\left(\frac{L R}{1.2\times 10^4 C(v,\rho)}\right)}-1\leq K.
\end{aligned}
\eeq
Combining \eqref{eq:qs+1<K1} with \eqref{eq:qs+1<K2}, we arrive at
\beq\label{eq:qs+1<K}
q_{s+1}\leq K.
\eeq

\subsubsection{Proof of part (a)}

By (ii) of \eqref{def:Ncase1} and Remark \ref{rmk:N2de}, we have 
\beq\nn
n\geq N\geq N_2(\omega, [a,b], \underline{L}(\omega), v, \rho)\geq N_2(\omega, [a,b], L(\omega, E), v, \rho).
\eeq
Hence by Lemma \ref{lm:un-uRn}, and \eqref{eq:betas<2cL}, we have,
\beq\label{eq:U1}
\begin{aligned}
\|\mathcal{U}_1(\cdot)\|_{L^\infty(\T)} 
\leq \frac{1}{100}L+\frac{1}{5}C({v, \rho})\frac{\log{q_{s+1}}}{q_s}
\leq &\frac{1}{100}L+\frac{2}{5}C({v, \rho})c(v,\rho)L
=\left(\frac{1}{100}+\frac{1}{9 \times 10^4}\right)L.
\end{aligned}
\eeq

By Proposition \ref{u2u3} and our choice of $\delta$, we have
\beq\label{eq:U2U3}
\begin{aligned}
\|\mathcal{U}_2(\cdot)+\mathcal{U}_3(\cdot)\|_{L^\infty(\T)}
\leq &\frac{\Lambda_v}{\delta n} \sup_{1\leq k\leq \delta^{-1}} \frac{1}{\|k\omega\|_{\T}}+4\pi \delta C(v,\rho)
\leq & \frac{1}{100}L,
\end{aligned}
\eeq
in which we used (iii) of \eqref{def:Ncase1}, see also \eqref{eq:part(a)N(c)},
\beq\nn
n\geq N\geq \frac{200\Lambda_v}{\delta \underline{L}} \sup_{1\leq k\leq \delta^{-1}} \frac{1}{\|k\omega\|_{\T}}
\geq  \frac{200\Lambda_v}{\delta L} \sup_{1\leq k\leq \delta^{-1}} \frac{1}{\|k\omega\|_{\T}}.
\eeq

Note that \eqref{eq:qs+1<K} verifies the condition $q_{s+1}\leq K$ of Proposition \ref{prop:u4u5u6}.
Hence Proposition \ref{prop:u4u5u6} implies that, 
\beq\nn
\begin{aligned}
\|\mathcal{U}_4(\cdot)+\mathcal{U}_5(\cdot)\|_{L^\infty(\T)}
\leq 120 C({v, \rho}) \left(\frac{\log{q_{s+1}}}{q_s}+\frac{\log K}{R}\right)
\leq 120 C({v, \rho}) \frac{\log{q_{s+1}}}{q_s}+\frac{1}{100}L.
\end{aligned}
\eeq
Taking \eqref{eq:betas<2cL} into account, we have
\beq\label{eq:U4U5}
\begin{aligned}
\|\mathcal{U}_4(\cdot)+\mathcal{U}_5(\cdot)\|_{L^\infty(\T)}
\leq 240 C({v, \rho})  c(v,\rho)L+\frac{1}{100}L= \frac{1}{60}L.
\end{aligned}
\eeq

Combining \eqref{eq:U1}, \eqref{eq:U2U3},  \eqref{eq:U4U5} with our choice of $c({v,\rho})$, see \eqref{def:ctildecThm(a)}, we have
\beq \label{eq:U12345}
\begin{aligned}
\|\sum_{j=1}^5 \mathcal{U}_j(\cdot)\|_{L^\infty(\T)} \leq \frac{1}{25}L.
\end{aligned}
\eeq

By \eqref{eq:u6} and \eqref{eq:qs+1<K2},
\beq\label{eq:U6}
\begin{aligned}
\|\mathcal{U}_6(\cdot)\|_{L^2(\T)}^2 \leq C^2({v, \rho})\frac{2}{K}
\leq &2C^2({v, \rho}) \exp{\left(-\frac{RL}{1.8 \times 10^4 C(v,\rho)}\right)}\\
< &2C^2({v, \rho}) \exp{\left(-\frac{nL}{10^7 (C_1+2)C({v, \rho})}\right)}.
\end{aligned}
\eeq

Combining \eqref{eq:U0}, \eqref{eq:U12345} with \eqref{eq:U6}, and using Markov's inequality, we obtain
\begin{align*}
{\rm mes}\left\lbrace x\in\T:\left| u_n(x)-L_n\right|>\frac{1}{20}L \right\rbrace
\leq &{\rm mes}\left\lbrace x\in\T:\left| \mathcal{U}_6(x) \right|>\frac{1}{100}L \right\rbrace\\
\leq &2\times10^4C^2({v, \rho}) L^{-2} \exp{\left(-\frac{n L}{10^7 (C_1+2)C({v, \rho})}\right)}\\
\leq &\exp{\left(-\frac{n L}{2\times 10^7(C_1+2)C({v, \rho})}\right)}\\
=&\exp{(-\tilde{c}(v,\rho)n L)},
\end{align*}
in which we used (iv) of \eqref{def:Ncase1},
\beq\nn
\begin{aligned}
n
\geq &2\times 10^7(C_1+2)C({v, \rho}) L^{-1} \log(2\times10^4C^2({v, \rho}) L^{-2}).
\end{aligned}
\eeq
This proves part (a) of Theorem \ref{thm:LDTmain}. \qed

\subsubsection{Parameters for part (b)}
In our decomposition of $u_n(x)-L_n$ in \eqref{eq:U0}, we choose parameters as follows:  
\beq\label{def:ctildecThm(b)}
\begin{aligned}
\delta &=\frac{1}{800\pi C({v, \rho})},\ \ \ \ \
R=\left[\frac{n L}{400 \Lambda_v}\right]+1,\\
K&=\left[\exp{\left(\frac{R L}{1.2\times 10^4 C({v, \rho})}\right)}\right],\ \ \ \ \
s=\max \{ s\in \N:\ q_s\leq R\}.
\end{aligned}
\eeq
It is clear that $q_s\leq R<q_{s+1}$.

Use the fact that $C_1>\Lambda_v$, see \eqref{def:C2}, and $\tilde{\underline{L}}\leq 1$, \eqref{def:Ncase1} implies
\beq\label{def:Ncase2}
n\geq 
\begin{cases}
(i').\ 400(\Lambda_v+1)q_{s_0}.\\
\\
(iii').\ \frac{200\Lambda_v}{\delta} \sup_{1\leq k\leq \delta^{-1}} \frac{1}{\|k\omega\|_{\T}},\\
\\
(iv').\ 2\times 10^7 \Lambda_v C(v, \rho) \log{(2\times 10^4 C^2(v, \rho) +e)}.
\end{cases}
\eeq

Note that (i') implies that
\beq
R>(400 \Lambda_v)^{-1} n L\geq (400 \Lambda_v)^{-1} n\geq q_{s_0}.
\eeq
By our definition of $s$, we have $s\geq s_0$. This, by \eqref{def:s0}, implies
\beq\label{eq:qs+1<K1'}
\begin{aligned}
q_{s+1}
\leq &\exp{\left(2c(v,\rho)\underline{L} q_s \right)}
\leq \exp{\left(2c(v,\rho){L} q_s \right)}\\
\leq &\exp{\left(2c(v,\rho) L R \right)}
\leq \exp{\left(\frac{L R}{1.8 \times 10^4 C(v,\rho)} \right)}.
\end{aligned}
\eeq
By (iv') of \eqref{def:Ncase2},
\beq\nn
n\geq 2 \times 10^7 \Lambda_v C(v, \rho) \log{(2\times 10^4 C^2(v, \rho) +e)}\geq 2\times 10^7 \Lambda_v C(v, \rho),
\eeq
hence 
\beq\nn
\begin{aligned}
\exp{\left(\frac{R L}{1.8 \times 10^4 C(v,\rho)} \right)}
\geq \exp{\left(\frac{n L^2}{7.2\times 10^6 \Lambda_v C(v,\rho)} \right)}
\geq &\exp{\left(\frac{n}{7.2\times 10^6 \Lambda_v C(v,\rho)} \right)}
>16.
\end{aligned}
\eeq
Thus, similar to \eqref{eq:qs+1<K2}, using the fact 
$x<x^{\frac{3}{2}}-1$, for $x>3$,
we have
\beq\label{eq:qs+1<K2'}
\exp{\left(\frac{R L}{1.8\times10^4 C(v,\rho)} \right)}\leq \exp{\left(\frac{R L}{1.2\times 10^4 C(v,\rho)} \right)}-1\leq K.
\eeq
Combining \eqref{eq:qs+1<K1'} with \eqref{eq:qs+1<K2'}, we obtain, similar to \eqref{eq:qs+1<K}, that
\beq\label{eq:qs+1<K'}
q_{s+1}\leq K.
\eeq

\subsubsection{Proof of part (b)}

We use the trivial upper bound in Lemma \ref{lm:un-uRn0} for $\mathcal{U}_1$,
\beq\label{eq:B1}
\|\mathcal{U}_1(\cdot)\|_{L^\infty(\T)} \leq 2\Lambda_v\frac{R}{n}\le \frac{1}{200}L+\frac{2\Lambda_v}{n}\leq \frac{1}{100}L,
\eeq
in which we used, see (i') of \eqref{def:Ncase2}, that
\beq\nn
n\geq 400 (\Lambda_v+1)q_{s_0}\geq 400 \Lambda_v L^{-1}.
\eeq

Proposition \ref{u2u3} yields that
\beq\label{eq:B2B3}
\begin{aligned}
\|\mathcal{U}_2(\cdot)+\mathcal{U}_3(\cdot)\|_{L^\infty(\T)}
\leq &\frac{\Lambda_v}{\delta n} \sup_{1\leq k\leq \delta^{-1}} \frac{1}{\|k\omega\|_{\T}}+
           4\pi C({v, \rho})\delta\\
\leq & \frac{1}{200}L+\frac{1}{200}L=\frac{1}{100}L.
\end{aligned}
\eeq
in which we used (iii') of \eqref{def:Ncase2},
\beq\label{eq:Nlb1'}
n
\geq \frac{200\Lambda_v}{\delta } \sup_{1\leq k\leq \delta^{-1}} \frac{1}{\|k\omega\|_{\T}}
\geq \frac{200\Lambda_v}{\delta L} \sup_{1\leq k\leq \delta^{-1}} \frac{1}{\|k\omega\|_{\T}}.
\eeq

Note that we have verified the condition $q_{s+1}\leq K$ in \eqref{eq:qs+1<K'}, Proposition \ref{eq:u4u5} implies that
\beq\nn
\begin{aligned}
\|\mathcal{U}_4(\cdot)+\mathcal{U}_5(\cdot)\|_{L^\infty(\T)}
\leq &120 C({v, \rho}) \left(\frac{\log{q_{s+1}}}{q_s}+\frac{\log{K}}{R}\right)\\
\leq &120 C({v, \rho}) \frac{\log{q_{s+1}}}{q_s}+\frac{1}{100}L.
\end{aligned}
\eeq
By \eqref{eq:betas<2cL}, we then have
\beq\label{eq:B4B5}
\begin{aligned}
\|\mathcal{U}_4(\cdot)+\mathcal{U}_5(\cdot)\|_{L^\infty(\T)}
\leq 240 C({v, \rho}) c(v,\rho) L+\frac{1}{100}L= \frac{1}{60}L.
\end{aligned}
\eeq

In view of $\mathcal{U}_6$, \eqref{eq:u6} and \eqref{eq:qs+1<K2'} yield that
\beq\label{eq:B6}
\begin{aligned}
\|\mathcal{U}_6(\cdot)\|_{L^2(\T)}^2 \leq C^2({v, \rho})\frac{2}{K}
\leq &2C^2({v, \rho}) \exp{\left(-\frac{R L}{1.8\times 10^4 C(v,\rho)} \right)}\\
\leq &2C^2({v, \rho}) \exp{\left(-\frac{nL^2}{10^7 \Lambda_v C({v, \rho})}\right)}.
\end{aligned}
\eeq

Combining \eqref{eq:B1},\eqref{eq:B2B3} \eqref{eq:B4B5} with \eqref{eq:B6}, we get that by Markov's inequality,
\begin{align*}
{\rm mes}\left\lbrace x\in\T:\left| u_n(x)-L_n\right|>\frac{1}{20}L \right\rbrace
\leq &{\rm mes}\left\lbrace x\in\T:\ \left| \mathcal{U}_6(x)\right|>\frac{1}{100}L \right\rbrace\\
\leq &2\times10^4 C^2({v, \rho}) L^{-2} \exp{\left(-\frac{nL^2}{10^7 \Lambda_v C({v, \rho})}\right)}\\
\leq &\exp{\left(-\frac{n L^2}{2\times 10^7 \Lambda_v C({v, \rho})}\right)},
\end{align*}
in which we used (iv') of \eqref{def:Ncase2}.
Using that $C_1>\Lambda_v$, we obtain 
\beq\nn
-\left(2\times 10^7 \Lambda_v C(v,\rho)\right)^{-1}<-\left(2\times 10^7 (C_1+2) C(v,\rho)\right)^{-1}=-\tilde{c}(v,\rho).
\eeq
Hence
\begin{align*}
{\rm mes}\left\lbrace x\in\T:\left| u_n(x)-L_n\right|>\frac{1}{20}L \right\rbrace
\leq \exp{\left(-\tilde{c}(v,\rho) nL^2\right)},
\end{align*}
as claimed.\qed

\subsection{Proof of Corollary \ref{cor:LDTlambda}}
In general, a large uniform norm of $v$  does not guarantee a positive Lyapunov exponent.
However if the potential function $v$ is of the form $\lambda f$, then the following well-known result by Sorets-Spencer \cite{SS} gives a lower bound of the Lyapunov exponent in the large coupling regime.

\begin{thm}\label{thm:SS}
For any non-constant real analytic potential $f$ with an analytic extension on $\{|\Im z|<\rho\}$, there exist constants $\lambda_0=\lambda_0(f)>0$ and $h_0=h_0(f)$ depending only on $f$, such that for all $E$, $\omega$ and $\lambda>\lambda_0$, the Lyapunov exponent $L(\omega,E)\geq \log{\lambda} + h_0.$
\end{thm}

Let $\lambda_0=\lambda_0(f)$ be given as in Theorem \ref{thm:SS}. For $\lambda>\lambda_1(f):=\max{(e^{-19 h_0}, 3, \lambda_0)}$ \footnote{$\lambda_0$ is in general large, however for some concrete examples, e.g. $f=\cos$, $\lambda_0=2$, cf~\cite{BJ}}, we have 
\begin{equation}\label{eq:SSlowerbd}
L(\omega, E)>\log{\lambda}+h_0>\frac{18}{19}\log{\lambda}> 1,
\end{equation}
holds uniformly in $\omega$ and $E$.
Let $\Lambda_v=\Lambda_{\lambda f}$ be defined as in \eqref{def:Cv}, we have
\beq\label{eq:D-1}
L(\omega, E)\leq \Lambda_{\lambda f}=\log{(3+2\lambda \|f\|_{L^{\infty}(\T)})}\leq \frac{20}{19}\log{\lambda},
\eeq
provided that $\lambda\geq \lambda_2(\|f\|_{L^\infty(\T)})$.

Let $C({\lambda f,\rho}), c({\lambda f,\rho})$ and $\tilde c({\lambda f,\rho})$ be defined as in (\ref{def:Cvrho}), \eqref{def:ctildecThm(a)}. 
With the help of Lemma \ref{lm:uniformsup-sup}, we can make the dependence of the three constants on $\lambda$ more explicit.

First, Lemma \ref{lm:uniformsup-sup} yields that there exists 
$C_0=C_0(f,\rho)$ such that 
\beq\label{eq:D-7}
C({\lambda f,\rho})\le C_0(f,\rho),
\eeq 
for any $\lambda\geq 0$.

Second, 
plugging \eqref{eq:D-1} and \eqref{eq:D-7} into our definition of $C_1$, see \eqref{def:C2}, we have,
\beq\label{eq:D-6}
C_1+2=4+\Lambda_{\lambda f}+ (8+4\pi C_3) C(\lambda f,\rho)\le 4+\frac{20}{19}\log{\lambda}+(8+4\pi C_3) C_0\le 2\log\lambda,
\eeq
provided that $\lambda\geq \lambda_3(f,\rho):= \max{(\lambda_2, \exp{\left(\frac{19}{18}(4+(8+4\pi C_3)C_0\right)} )}$. 
Thus putting \eqref{eq:D-7} and \eqref{eq:D-6} together, we have that for $\lambda\geq \lambda_3$,
\beq\label{eq:D-8}
\tilde{c}(\lambda f,\rho)=\big(2\times 10^7(C_1+2)C({v, \rho})\big)^{-1}\geq \big(4 \times 10^7 C_0 \log{\lambda}\big)^{-1}.
\eeq

Third, note that \eqref{eq:D-7} also yields
\beq\label{eq:D-5}
c(\lambda f, \rho)=(36000 C(\lambda f,\rho))^{-1}\geq (36000 C_0)^{-1}.
\eeq

Let us take 
\begin{align*}
\tilde{\lambda}(f,\rho):=\max{(\lambda_1, \lambda_3)},
\end{align*} 
and $\lambda>\tilde{\lambda}$.
We are in the place to apply Theorem \ref{thm:LDTmain}.
Let us note that by \eqref{eq:SSlowerbd}, we always have $L(\omega, E)>1$, hence we will only apply part (b).
One condition of the theorem is $0\leq \beta(\omega)<c(\lambda f, \rho)L(\omega, E)$.
In view of \eqref{eq:D-5} and $L(\omega, E)>\frac{18}{19}\log{\lambda}$, this condition will always be satisfied if
\beq\label{eq:D-9}
\beta(\omega)<(36000 C_0)^{-1} \frac{18}{19} \log{\lambda}=(38000 C_0)^{-1}\log{\lambda}=: B^{-1}\log{\lambda}.
\eeq
Therefore, for $\lambda> \max{(\tilde{\lambda}, \exp{(B \beta(\omega)}))}$, part (b) of Theorem \eqref{thm:LDTmain} implies
\beq\label{eq1:Cor15}
{\rm mes}\left\lbrace x\in\T:\mid u_n(\omega, E; x)-L_n(\omega, E)\mid>\frac{1}{20}L(\omega,E) \right\rbrace\\
\leq \exp{\left(-\tilde{c}({\lambda f, \rho}) L^2(\omega, E)n\right)}.
\eeq
Using upper and lower bounds of $L(\omega,E)$ in \eqref{eq:D-1} and \eqref{eq:SSlowerbd}, we obtain from \eqref{eq1:Cor15} that
\beq\label{eq:D-2}
\begin{aligned}
&{\rm mes}\left\lbrace x\in\T:\mid u_n(\omega, E; x)-L_n(\omega, E)\mid>\frac{1}{19}\log{\lambda} \right\rbrace\\
&\qquad\qquad \leq {\rm mes}\left\lbrace x\in\T:\mid u_n(\omega, E; x)-L_n(\omega, E)\mid>\frac{1}{20}L(\omega,E) \right\rbrace\\
&\qquad\qquad \leq \exp{\left(-\tilde{c}({\lambda f, \rho}) L^2(\omega, E)n\right)}\\
&\qquad\qquad \leq \exp{\left(-\tilde{c}({\lambda f, \rho}) \frac{18^2 (\log{\lambda})^2}{19^2}\,n\right)}\\
&\qquad\qquad \leq \exp{\left(-n\frac{\log{\lambda}}{5 \times 10^7 C_0}\right)}=:\exp{(-n b \log{\lambda})},
\end{aligned}
\eeq
in which we used $\eqref{eq:D-8}$ in the last inequality. \qed

\section{The proofs of the lemmas}\label{sec:proofoflemma}

\subsection{Proof of Lemma \ref{lm:ukold}}\label{sec:proofukold}
We need the following result. 
\begin{lm}\label{lm:sup-supGS08}\cite[Lemma 2.2]{GS2}
Let $u:\Omega\to \R$ be a subharmonic function on a domain $\Omega\subset\C$.
Suppose that $\partial \Omega$ consists of finitely many piece-wise $C^1$ curves.
There exists a positive measure $\mu$ on~$\Omega$ such that for any $\Omega_1\Subset \Omega$
(i.e., $\Omega_1$ is a compactly contained sub-region of~$\Omega$)
\begin{equation}
\label{eq:GS08decomp}
u(z) = \int_{\Omega_1} \log|z-\zeta|\,d\mu(\zeta) + h(z)
\end{equation}
where $h$ is harmonic on~$\Omega_1$ and $\mu$ is unique with this property.
Moreover, $\mu$ and $h$ satisfy the bounds
\begin{eqnarray}
\mu(\Omega_1) &\le& C(\Omega,\Omega_1)\,(\sup_{\Omega} u - \sup_{\Omega_1} u) \label{eq:mubound} \label{eq:GS08mu}\\
\|h-\sup_{\Omega_1}u\|_{L^\infty(\Omega_2)} &\le& C(\Omega,\Omega_1,\Omega_2)\,(\sup_{\Omega} u - \sup_{\Omega_1} u)
\label{eq:GS08h}
\end{eqnarray}
for any $\Omega_2\Subset\Omega_1$.
\end{lm}

Note that $u_n(z)$ is a bounded subharmonic function on
$\Omega:=\{z:|{\rm Re}z|<1,\ |{\rm Im}z|<\rho\}$. We consider the following nested domains $\Omega_0\Subset\Omega_2\Subset\Omega_1\Subset\Omega$, where
\beq\label{def:Omega012}
\begin{aligned}
\Omega_1=&\{z: |{\rm Re}z|\le\frac{5}{6},\
                           |{\rm Im}z|<\frac{\rho}{2}\} \\
\Omega_2=&\{z: |{\rm Re}z|\le\frac{4}{5},\
                         |{\rm Im}z|<\frac{\rho}{4}\} \\
\Omega_0=&\{z: |{\rm Re}z|\le\frac{3}{4},\ |{\rm Im}z|=0\}
=\left[-\frac{3}{4},\frac{3}{4}\right].
\end{aligned}
\eeq
Now we apply Lemma \ref{lm:sup-supGS08} to $u(z)=u_n(z)$ on $\Omega$. We have then a positive measure $\mu$ and a harmonic function $h$ on $\Omega_1$ satisfying (\ref{eq:GS08decomp}), (\ref{eq:GS08mu}) and (\ref{eq:GS08h}). 

Since $h-\sup_{\Omega_1}u$ is a harmonic function, by the Poission integral formula and  (\ref{eq:GS08h}), we have
\begin{equation}\label{eq:GS08h'}
\max{(\|\partial_x h\|_{L^\infty(\Omega_0)}, \|\partial^2_x h\|_{L^\infty(\Omega_0)})} \le C(\Omega, \Omega_1, \Omega_2, \Omega_0)(\sup_{\Omega} u - \sup_{\Omega_1} u).
\end{equation}
We only need to the bound for $\partial_x h$ here, we will use the one for $\partial^2_x h$ in Sec.~\ref{sec:uknew}.

Combine (\ref{eq:GS08decomp}) with the technique in \cite{BG}, one can then that for some absolute constant $C_2>0$, the following holds for any $k\neq 0$:
\begin{equation}\label{eq:GS08uk}
\left| \hat u_n(k) \right| \le \frac{C_2}{|k|} \left(\mu(\Omega_1)+\|\partial_x h\|_{L^\infty(\Omega_0)}
+\|h-\sup_{\Omega_1}u_n\|_{L^\infty(\Omega_0)} \right).
\end{equation}
Clearly, (\ref{eq:ukold}),(\ref{def:Cvrho}) follow directly from (\ref{eq:GS08mu})-(\ref{eq:GS08uk}) by setting 
\beq
\alpha(\rho):=C_2 \max{(C(\Omega, \Omega_1), C(\Omega, \Omega_1, \Omega_2), C(\Omega, \Omega_1, \Omega_2, \Omega_0))}.
\eeq
This finishes the proof of Lemma \ref{lm:ukold}.
We will include the proof of \eqref{eq:GS08uk} in Appendix \ref{sec:GS08uk}.
\qed

\subsection{Proof of Lemma \ref{lm:uniformsup-sup}}\label{sec:uniformsup-sup}
On one hand, for any $E\in{\cal N}$, trivially we have
$$\sup_{|{\rm Im}z|<\rho}\|A_j(E,z)\|\le 2\lambda \|f\|_\rho+2\le 3\lambda \|f\|_\rho,\ \textrm{provided} \ \lambda>2\|f\|^{-1}_\rho $$
and $$\sup_{|{\rm Im}z|<\rho }u_n(z)\le \log\Big(3\lambda \|f\|_\rho\Big)$$
On the other hand, since $f$ is non-constant analytic on $|{\rm Im}z|<\rho$, for $\delta=\rho/2$,
there exists $\varepsilon_0=\varepsilon_0(f)>0$ such that
$$\inf_{E_1}\sup_{y\in(\delta/2,\delta)}\inf_{x}|f(x+ {i} y)-E_1|>\varepsilon_0$$
This implies that  for any $\lambda, E$, there is $y_0\in (\delta/2,\delta)$ such that $\forall x$
$$|f(x+ {i} y_0)-{E}/{\lambda}|>\varepsilon_0$$
The computation contained in \cite[Appendix]{BG} shows that for $\lambda>2\varepsilon_0^{-1}$,
\begin{eqnarray}
  \|M_n( {i} y_0,E)\| 
   \ge  \prod_{j=1}^n\Big(|\lambda f(j\omega+ {i} y_0)-E|-1\Big)
   \ge  \Big(\lambda \epsilon_0-1\Big)^n 
   \ge  \Big(\frac{1}{2}\lambda \epsilon_0\Big)^n.
\end{eqnarray}
Therefore,
$$\sup_{|{\rm Im}z|<\rho/2 }u_n(z)\ge u_n( {i} y_0)=\frac{1}{n}\log \|M_n( {i} y_0,E)\|\ge \log\left(\frac{1}{2}\lambda \varepsilon_0\right) $$
Clearly, we have that for $\lambda>\max\{2\|f\|^{-1}_\rho,3\varepsilon_0^{-1}\}$,
$$ \sup_{|{\rm Im}z|<\rho }u_n(z)-\sup_{|{\rm Im}z|<\rho/2 }u_n(z)
\le \log\big(3\lambda \|f\|_\rho\big)-\log\left(\frac{1}{2}\lambda \varepsilon_0\right)=\log\left(\frac{6\|f\|_\rho}{\varepsilon_0}\right)$$
Therefore by \eqref{def:Cvrho},
$$C({\lambda f,\rho})\le \alpha(\rho) \log\left(\frac{6\|f\|_\rho}{\varepsilon_0}\right)=: C_0(f, \rho)\ \
\textrm{independent of} \ \ \lambda,$$
as desired.
\qed

\subsection{Proof of Lemma \ref{lm:uknew}}\label{sec:uknew}
We have that by \eqref{trivialbdd},
\beq\label{eq:ux+omega-ux}
\begin{aligned}
  &\| u_n(\cdot+\omega)-u_n(\cdot)\| _{L^{\infty}(\T)}\\
=&\frac{1}{n}\lv \log{\|M_n(\cdot+\omega)\|}- \log{\|M_n(\cdot)\|} \rv_{L^{\infty}(\T)}\\
\leq &\frac{1}{n}\lv \log{\|M_1(\cdot+n\omega)\|}+\log{\|M_{n-1}(\cdot+\omega)\|}+\log{\|M_1(\cdot)\|}-\log{\|M_{n-1}(\cdot+\omega)\|} \rv_{L^{\infty}(\T)}\\
\leq &\frac{2\Lambda_v}{n}.
\end{aligned}
\eeq
This implies
\beq\label{eq:hatuk-hatuk}
\begin{aligned}
  \lvert \hat{u}_n(k)e^{2\pi i k\omega}-\hat{u}_n(k) \rvert
=&\lvert \int_{T}u_n(x+\omega) e^{-2\pi i kx}\ \mathrm{d}x-\int_{\T} u(x)e^{-2\pi i x}\ \mathrm{d}x\rvert\\
\leq &\| u_n(\cdot+\omega)-u_n(\cdot)\| _{L^{\infty}(\T)}
\leq \frac{2\Lambda_v}{n}
\end{aligned}
\eeq
\eqref{eq:hatuk-hatuk} implies
\begin{align*}
2\lvert \hat{u}_n(k)\rvert \sin{(\pi\|k\omega\|_{\T})} \leq \frac{2\Lambda_v}{n},
\end{align*}
hence by $\sin{(\pi x)}\geq 2x$ for $0\leq x\leq \frac{1}{2}$, we get that for $k\neq 0$,
\begin{align*}
\lvert \hat{u}_n(k)\rvert\leq \frac{\Lambda_v}{2n\|k\omega\|_{\T}},
\end{align*}
as stated. \qed

Before we move on, let us mention a simple consequence of \eqref{eq:ux+omega-ux}:
\beq\label{eq:ux+jomega-ux}
\| u_n(\cdot+\omega)-u_n(\cdot)\| _{L^{\infty}(\T)}\leq \frac{2\Lambda_v |j|}{n},
\eeq
this estimate will be used in several parts of the argument.

\subsection{Proof of Lemma \ref{lm:uRupper}}
Let $R\geq R_0(L)$ and $n\geq N_3(v,\omega, L)$, where
\beq\label{def:R0}
R_0:=144 L^{-5},
\eeq
and 
\beq\label{def:N3}
N_3:=2\Lambda_v L^{-2} \sup_{1\leq |k|\leq L^{-1}}\frac{1}{\|k\omega\|_{\T}}.
\eeq

Lemma \ref{lm:sup-supGS08} implies that $u_n$ has a Riesz-representation with a positive measure $\mu$ and a harmonic function $h$.
Let us take 
\beq\label{def:deltauRupper}
\delta=(L R)^{-1},
\eeq 
and
\beq\label{def:udelta}
u_{n,\delta}(x)=\int_{\Omega_1} \log{(|x-w|+\delta)}\ \mu(\mathrm{d} w)+h(x),
\eeq
We then have, point-wisely,
\beq\label{eq:u<udelta}
u_n(x)\leq u_{n,\delta}(x).
\eeq
It is clear from our definitions of $R_0$ and $\delta$ that, 
\beq\label{eq:delta<}
\delta\leq \frac{L^4}{144}<\frac{1}{144}.
\eeq

\subsubsection{Fourier coefficients decay for $u_{n,\delta}$}
The following two inequalities \eqref{eq:udeltak} and \eqref{eq:udeltak3} are (2.4) and (2.3) of \cite{BG} (see also (8.12) of \cite{B1}).
We include their proofs in Appendix \ref{sec:undeltaproof}.
\begin{lm}\label{lm:Fourierundelta}
Let $C(v,\rho)$ be defined as in \eqref{def:Cvrho}.
There exists an absolute constant $C_3$ such that for any $k\in \Z$, we have
\beq\label{eq:udeltak}
\lvert \hat{u}_{n,\delta}(k)\rvert\leq \lvert \hat{u}_n(k)\rvert +3\delta \log{\delta^{-1}},
\eeq
and for any $k\neq 0$,
\beq\label{eq:udeltak3}
\lvert \hat{u}_{n,\delta}(k)\rvert\leq C_3 C({v, \rho})\min\left(\frac{1}{|k|}, \frac{1}{k^2\delta}\right),
\eeq
holds for $k\neq 0$.
\end{lm}
Note that \eqref{eq:udeltak} together with Lemma \ref{lm:uknew} leads to the following corollary.
\begin{cor}
For $k\neq 0$, we have
\beq\label{eq:udeltak2}
\lvert \hat{u}_{n,\delta}(k)\rvert\leq \frac{\Lambda_v}{2n\|k\omega\|_{\T}}+3\delta\log{\delta^{-1}}.
\eeq
\end{cor}

\subsubsection{Proof of Lemma \ref{lm:uRupper}}
Let $s\in \N$ be such that $q_s\leq R<q_{s+1}$.
Recall that our definition of $u^{(R)}$, see \eqref{def:uR}.
\eqref{eq:u<udelta} clearly yields
\beq\nn
\begin{aligned}
0\leq u_n^{(R)}(x)
\leq u_{n,\delta}^{(R)}(x)
\end{aligned}
\eeq
Let $F_R(k)$ be as in \eqref{def:FRk}, invoking \eqref{eq:unRFR}, we have
\beq\label{eq:A1}
\begin{aligned}
0\leq u_n^{(R)}(x)
\leq u_{n,\delta}^{(R)}(x)
=\hat{u}_{n,\delta}(0)+\sum_{k\neq 0}\hat{u}_{n,\delta}(k) F_R(k).
\end{aligned}
\eeq
We now split the Fourier series in \ref{eq:A1} into low/high-frequency parts,
\beq\label{eq:A2}
\begin{aligned}
u_{n,\delta}^{(R)}(x)
= &\hat{u}_{n,\delta}(0)+\sum_{1\leq |k|\leq {q_{s+1}}/{4}}\hat{u}_{n,\delta}(k) F_R(k)+\sum_{|k|>{q_{s+1}}/{4}}\hat{u}_{n,\delta}(k) F_R(k)\\
=:&\hat{u}_{n,\delta}(0)+\mathcal{S}_1+\mathcal{S}_2.
\end{aligned}
\eeq

Using the $(k^2\delta)^{-1}$ bound of $|\hat{u}_{n,\delta}(k)|$ in \eqref{eq:udeltak3} and $|F_R(k)|\leq 1$ in \eqref{eq:FRkleq}, we have
\beq\label{eq:S2}
|\mathcal{S}_2|
\leq \sum_{|k|>{q_{s+1}}/{4}}\lvert \hat{u}_{n,\delta}(k)\rvert 
\leq \sum_{|k|>{q_{s+1}}/{4}}\frac{C({v, \rho})}{k^2\delta}
\leq \frac{8C({v, \rho})}{q_{s+1}\delta}
\leq \frac{8C({v, \rho})}{\delta R}
= 8C({v, \rho})L,
\eeq
in which we used $R<q_{s+1}$ and our choice of $\delta$, see \eqref{def:deltauRupper}.

We further decompose $\mathcal{S}_1$ into
\beq\label{eq:S1}
\begin{aligned}
|\mathcal{S}_1|
&\leq \left( \sum_{1\leq |k|\leq L^{-1}}+\sum_{L^{-1}<|k|< q_s/4}+\sum_{q_s/4\leq |k|\leq q_{s+1}/4}\right) \lvert \hat{u}_{n,\delta}(k)\rvert F_R(k)\\
&=:\mathcal{S}_{1,1}+\mathcal{S}_{1,2}+\mathcal{S}_{1,3}.
\end{aligned}
\eeq

By \eqref{eq:udeltak2} and $|F_R(k)|\leq 1$, see \eqref{eq:FRkleq}, we have
\beq\nn
\begin{aligned}
\mathcal{S}_{1,1}
\leq &\sum_{1\leq |k|\leq L^{-1}} \left(\frac{\Lambda_v}{2n\|k\omega\|_{\T}}+3\delta\log{\delta^{-1}}\right)\\
\leq &\frac{2}{L} \left(\frac{\Lambda_v}{2n}\sup_{1\leq |k|\leq L^{-1}} \frac{1}{\|k\omega\|_{\T}}+\frac{3}{R L}\log{(RL)}\right).
\end{aligned}
\eeq
Using a trivial estimate: $\log{x}\leq \sqrt{x}$ that holds for any $x>0$, we obtain
\beq\label{eq:S11}
\begin{aligned}
\mathcal{S}_{1,1}
\leq \left(\frac{\Lambda_v}{nL}\sup_{1\leq |k|\leq L^{-1}} \frac{1}{\|k\omega\|_{\T}}+\frac{6}{\sqrt{R L^3}}\right)\leq L,
\end{aligned}
\eeq
in the last step we used $R\geq R_0= 144 L^{-5}$ and $n\geq N_3$, see \eqref{def:R0} and \eqref{def:N3}.

Using the $|k|^{-1}$ bound of $|\hat{u}_{n,\delta}(k)|$ in \eqref{eq:udeltak3}, and non-trivial bound of $|F_R(k)|$ in \eqref{eq:FRkleq}, we have
\beq\nn
\begin{aligned}
\mathcal{S}_{1,2}
\leq &2C_3 C(v,\rho) L \sum_{L^{-1}< |k|<q_s/4}\frac{1}{1+R^2\|k\omega\|_{\T}^2}
\leq &2C_3 C(v,\rho) L \sum_{1\leq |k|<q_s/4}\frac{1}{1+R^2\|k\omega\|_{\T}^2}.
\end{aligned}
\eeq
Applying \eqref{eq:lowqsR}, we obtain
\beq\label{eq:S12}
\begin{aligned}
\mathcal{S}_{1,2}\leq 4\pi C_3 C(v,\rho) L \frac{q_s}{R}\leq 4\pi C_3 C(v,\rho) L,
\end{aligned}
\eeq
in which we used $q_s\leq R$.

The estimate of $\mathcal{S}_{1,3}$ is similar to that of $\mathcal{S}_{1,2}$, except that we use \eqref{eq:low1+qsR} instead of \eqref{eq:lowqsR}.
Indeed, by \ref{eq:udeltak3}, \ref{eq:FRkleq}, \eqref{eq:low1+qsR}, we have
\beq\label{eq:S13}
\begin{aligned}
\mathcal{S}_{1,3}
\leq &\sum_{\ell=1}^{[q_{s+1}/q_s]+1} \sum_{|k|\in [\ell q_s/4, (\ell+1)q_s/4)} |\hat{u}_{n,\delta}(k) F_R(k)|\\
\leq &\sum_{\ell=1}^{[q_{s+1}/q_s]+1} \sum_{|k|\in [\ell q_s/4, (\ell+1)q_s/4)} \frac{2 \lvert \hat{u}_{n,\delta}(k)\rvert}{1+R^2\|k\omega\|_{\T}^2}\\
\leq &\sum_{\ell=1}^{[q_{s+1}/q_s]+1} \frac{8 C({v, \rho})}{\ell q_s} \left(2+4\pi \frac{q_s}{R}\right)\\
\leq &120 C(v,\rho) \frac{\log{q_{s+1}}}{q_s}.
\end{aligned}
\eeq

Note that by \eqref{eq:udeltak} with $k=0$, we have
\beq\nn
\begin{aligned}
\hat{u}_{n,\delta}(0)
\leq L_n+\frac{1}{RL}\log{(RL)}.
\end{aligned}
\eeq
Trivial estimate $\log x\leq \sqrt{x}$ for $x>0$ implies
\beq\label{eq:S0}
\begin{aligned}
\hat{u}_{n,\delta}(0)
\leq L_n+\frac{1}{\sqrt{R L}}
\leq L_n+\frac{L^2}{12}
<L_n+L,
\end{aligned}
\eeq
in which we used $R\geq R_0\geq 144 L^{-5}$ and $0<L<1$.

Combining \eqref{eq:A1}, \eqref{eq:A2}, \eqref{eq:S2}, \eqref{eq:S1}, \eqref{eq:S11}, \eqref{eq:S12}, \eqref{eq:S13} with \eqref{eq:S0}, we arrive at
\beq\nn
0\leq u_n^{(R)}(x)\leq L_n+(2+8 C(v, \rho) +4\pi C_3 C(v,\rho)) L+120 C(v, \rho) \frac{\log{q_{s+1}}}{q_s},
\eeq
holds uniformly in $x$.
\qed

\subsection{Proof of Lemma \ref{lm:unupper}}
We apply Lemma \ref{lm:uRupper} to $R=\lfloor 3Ln \rfloor$.
The conditions $R\geq R_0$ and $n\geq N_3$, see \eqref{def:R0} and \eqref{def:N3}, can be reduced to
\beq
\begin{aligned}
n\geq N_0(\omega, L, v, \rho):=L^{-2} \max\left( 2\Lambda_v \sup_{1\leq |k|\leq L^{-1}}\frac{1}{\|k\omega\|_{\T}},\ 49 L^{-4}\right).
\end{aligned}
\eeq
Indeed, due to $0<L<1$, we have
$$
R\geq 3Ln-1\geq 147 L^{-5}-1>144 L^{-5}.
$$

Now for $n\geq N_0$, Lemma \ref{lm:uRupper} implies
\beq\label{eq:Q1}
\begin{aligned}
0\leq u_n(x)
\leq &\lvert u_n(x)-u_n^{(R)}(x)\rvert+ u_n^{(R)}(x)\\
\leq &\lvert u_n(x)-u_n^{(R)}(x)\rvert+L_n+(2+8 C(v, \rho) +4\pi C_3 C(v,\rho)) L+120 C(v, \rho) \frac{\log{q_{s+1}}}{q_s}.
\end{aligned}
\eeq
By \eqref{eq:ux+omega-ux}, we have
\beq\label{eq:Q2}
\lvert u_n(x)-u_n^{(R)}(x)\rvert\leq \sum_{|j|<R}\frac{R-|j|}{R^2}\cdot  \frac{2 \Lambda_v |j|}{n}=\frac{(R^2-1)\Lambda_v}{3Rn}<\Lambda_v L.
\eeq
Hence combining \eqref{eq:Q1} with \eqref{eq:Q2}, we get
\begin{align*}
0\leq u_n(x) \leq L_n+(2+\Lambda_v+8 C(v, \rho) +4\pi C_3 C(v,\rho)) L+120 C(v, \rho) \frac{\log{q_{s+1}}}{q_s},
\end{align*}
holds uniformly in $x$.
\qed


\subsection{Proof of Lemma \ref{lm:un-uRn}}
Let 
\beq\nn
N_2=\max{(150 \Lambda_v N_1 L^{-1},\ 400 (C_1+2) N_1+1 )},
\eeq
be as in \eqref{def:N2}.
Let $n\geq N_2$ and $R=\lfloor (400(C_1+2))^{-1} n\rfloor +1$.

By \eqref{eq:ux+jomega-ux}, we have
\beq\label{eq:Q0}
\begin{aligned}
       \lv u_n(\cdot)-u_n^{(R)}(\cdot)\rv_{L^\infty(\T)}
\leq &\sum_{|j|<R}\frac{R-|j|}{R^2} \lv u_n(\cdot)-u_n(\cdot+j\omega) \rv_{L^\infty(\T)}\\
\leq &\sum_{|j|<R}\frac{|j|(R-|j|)}{n R^2}\lv u_j(\cdot+n\omega)+u_j(\cdot) \rv_{L^{\infty}(\T)}\\
\leq &\sum_{|j|<R}\frac{2|j|(R-|j|)}{n R^2}\lv u_j(\cdot) \rv_{L^{\infty}(\T)}.
\end{aligned}
\eeq
By our choice of $R$ and $n\geq N_2\geq 400(C_1+2)N_1+1$, we have
\beq
R\geq \frac{n}{400(C_1+2)}>N_1.
\eeq
We could split the sum in \eqref{eq:Q0} into:
\beq\label{eq:Q3}
\begin{aligned}
       &\lv u_n(\cdot)-u_n^{(R)}(\cdot)\rv_{L^\infty(\T)}\\
\leq &\sum_{|j|< N_1}\frac{2|j|(R-|j|)}{n R^2}\lv u_j(\cdot) \rv_{L^{\infty}(\T)}+\sum_{N_1\leq |j|<R}\frac{2|j|(R-|j|)}{n R^2}\lv u_j(\cdot) \rv_{L^{\infty}(\T)}.
\end{aligned}
\eeq
We will use trivial upper bound $\|u_j(\cdot)\|_{L^\infty(\T)}\leq \Lambda_v$, see \eqref{trivialbdd}, in the first summation of \eqref{eq:Q3}. 
Note that $j\geq N_1\geq N_0$, hence we can apply Lemma \ref{lm:unupper} to $u_j$ in the second sum.
We have
\beq\label{eq:Q4x}
\begin{aligned}
&\lv u_n(\cdot)-u_n^{(R)}(\cdot)\rv_{L^\infty(\T)}\\
\leq &\sum_{|j|< N_1}\frac{2\Lambda_v |j|(R-|j|)}{n R^2}+\sum_{N_1\leq |j|<R}\frac{2|j|(R-|j|)}{n R^2} \Big(L_j+C_1 L+120 C(v, \rho) \frac{\log{q_{s+1}}}{q_s}\Big).
\end{aligned}
\eeq
For $j\geq N_1\geq \widetilde{N}_0+1$, Lemma \ref{lm:uniformtildeN0} implies $L_j\leq 21 L/20<2L$, hence 
\beq\label{eq:Q4y}
\begin{aligned}
&\lv u_n(\cdot)-u_n^{(R)}(\cdot)\rv_{L^\infty(\T)}\\
\leq &\sum_{|j|< N_1}\frac{2\Lambda_v |j|(R-|j|)}{n R^2}+\sum_{N_1\leq |j|<R}\frac{2|j|(R-|j|)}{n R^2} \Big((C_1+1) L+120 C(v, \rho) \frac{\log{q_{s+1}}}{q_s}\Big).
\end{aligned}
\eeq
Use that 
\beq\nn
\begin{aligned}
\sum_{|j|<N_1}\frac{2|j| (R-|j|)}{R^2}=N_1 \frac{2(N_1-1) (3R-2N_1+1) }{3R^2}\leq \frac{3}{4}N_1,
\end{aligned}
\eeq
and
\beq\nn
\begin{aligned}
\sum_{N_1\leq |j|<R}\frac{2|j| (R-|j|)}{R^2}
=(R+1-N_1) \frac{2\left(R(R-1)+(R+1)N_1-2N_1^2\right)}{3R^2}
\leq R\frac{2(R^2-R)}{3R^2}=
\frac{2}{3}(R-1).
\end{aligned}
\eeq
We could control \eqref{eq:Q4y} by
\beq\label{eq:Q4w}
\begin{aligned}
\lv u_n(\cdot)-u_n^{(R)}(\cdot)\rv_{L^\infty(\T)} \leq \frac{3 \Lambda_v N_1}{4 n}+\frac{2(R-1)}{3n} \Big((C_1+2) L+120 C(v, \rho) \frac{\log{q_{s+1}}}{q_s}\Big).
\end{aligned}
\eeq
For the first term in \eqref{eq:Q4w}, note that
$n\geq N_2\geq 150 \Lambda_v N_1 L^{-1}$ implies
\beq\label{eq:Q4z}
\begin{aligned}
\frac{3 \Lambda_v N_1}{4 n}\leq  \frac{1}{200}L.
\end{aligned}
\eeq
For the second term, we plug in $R=\lfloor 400^{-1} (C_1+2)^{-1} n \rfloor+1$, then we have
\beq\label{eq:Q4r}
\begin{aligned}
&\frac{2\left(C_1+2\right) (R-1)}{3 n}L< \frac{1}{200}L,\ \ \  \text{and}\\
&\frac{2(R-1)}{3 n}\cdot 120 C(v, \rho) \frac{\log{q_{s+1}}}{q_s}\leq \frac{4 C(v,\rho)}{15 (C_1+2)} \frac{\log{q_{s+1}}}{q_s}\leq \frac{1}{5} C(v, \rho) \frac{\log{q_{s+1}}}{q_s}.
\end{aligned}
\eeq
Incorporating the estimates in \eqref{eq:Q4z} and \eqref{eq:Q4r} into \eqref{eq:Q4w}, we have
\beq\nn
\lv u_n(\cdot)-u_n^{(R)}(\cdot)\rv_{L^\infty(\T)} \leq \frac{1}{100}L +\frac{1}{5} C(v,\rho) \frac{\log{q_{s+1}}}{q_s},
\eeq
as stated.
\qed

\section{Refined H\"older continuity}\label{sec:proofofHolder}

H\"older regularity of $L(E)$ follows from combing LDT with AP. This scheme was developed by Goldstein and Schlag in \cite{GS1}, and has shown to be not restricted to quasi-periodic cocycles, see e.g. \cite{BGS} for skew-shift. 
This scheme was extended to general cocycles in any dimension, in a recent monograph \cite{DKbook}.
Recently, it was also used to study the  1-d Anderson model in\cite{BDFGVWZ}. We sketch the proof below in our setting, making the indenpendence of the H\"older exponent explicit. 
\subsection{Proof of Theorem \ref{thm:holder1}}
Fix $(\omega_0,E_0)\in (\R\setminus \Q)\times{\cal N}_v$ with $L(\omega_0,E_0)=\gamma>0$. 
As we explained in Remark \ref{rmk:QIexists} that the neighborhood $U \times I$ as in Theorem \ref{thm:holder1} always exists.
For any $(\omega, E)\in U\times I$:
\begin{equation}\label{eq:lypgamma}
 \frac{18}{19}\gamma\leq L(\omega,E)\leq \frac{20}{19}\gamma. 
\end{equation}

Let $c({v, \rho})$ and $\tilde{c}=\tilde{c}({v, \rho})$ be the constants in Theorem \ref{thm:LDTmain}.
Define a subset $\tilde{U}$ of $U$ as follows
\beq\label{def:tildeQ}
\tilde{U}:=\{\omega\in \R\setminus \Q:\ 0\leq \beta(\omega)< c(v,\rho)\gamma/2\}\cap U.
\eeq
In particular, $\tilde{U}$ contains all the Diophantine numbers in $U$, thus $\mathrm{mes}(U\setminus \tilde{U})=0$.

We are going to apply Theorem \ref{thm:LDTmain} on interval $[a,b]=I$. 
Note that for any $\omega\in \tilde{U}$, by (\ref{eq:lypgamma}), we have
\begin{equation}\label{eq:betaLDTcondition}
0\leq \beta(\omega)<\frac{1}{2}c(v,\rho)\gamma<c({v,\rho})\inf_{E\in I} L(\omega,E).
\end{equation}
Hence the condition of Theorem \ref{thm:LDTmain} is verified.
Let $N=N(\omega, \inf_{E\in I} L(\omega, E), v, \rho)$ be as in \eqref{def:Ncase1}, which is the constant in Theorem \ref{thm:LDTmain}.
Let $\tilde{N}=N(\omega, \frac{18}{19}\gamma, v, \rho)$ be the constant defined in \eqref{def:Ncase1} with $\underline{L}=\frac{18}{19}\gamma$.
Then by \eqref{eq:lypgamma} and Remark \ref{rmk:Nde}, we have $\tilde{N}\geq N$.
Let $$\Omega_n(\omega,E):=\left\lbrace x\in\T:\mid u_n(\omega, E; x)-L_n(\omega, E)\mid>\frac{1}{20}L(\omega, E) \right\rbrace. $$

Theorem \ref{thm:LDTmain} implies that for $n\geq \tilde{N}\geq N$ and any $(\omega,E)\in \tilde{U} \times I$, we have
\begin{equation}\label{eq:LDTgamma}
  {\rm mes} \left(\Omega_n(\omega,E)\right) \leq e^{-\tilde{c}nL(\omega, E)}
   \leq e^{-\tilde{c} n\gamma/2 },
\end{equation}
in which we used $L(\omega, E)\geq \frac{18}{19}\gamma>\frac{1}{2}\gamma$, see \eqref{eq:lypgamma}.

In the rest of the section, we will fix $\omega \in \tilde{U}$ and denote $L(E)=L(\omega,E)$, $L_n(E)=L_n(\omega,E)$ for simplicity whenever it is clear. 
Apply Lemma \ref{lm:uniformtildeN0} to the interval $I$. Let $\widetilde{N}_0(\omega, I, v)$ be given as in Lemma \ref{lm:uniformtildeN0}. Then for any $n>\widetilde{N}_0$ and $E\in I$, we have
\begin{equation}\label{eq:uniformLn}
L(E)\le L_n(E)<\left(1+\frac{1}{20}\right)L(E).
\end{equation}
Combining \eqref{eq:uniformLn} with the fact that $L_{2n}(E)\leq L_n(E)$, we have for all $n>\widetilde{N}_0$ and $E\in I$,
\begin{equation}\label{eq:LnL2n}
 0\le L_n(E)-L_{2n}(E)<\frac{1}{20}L(E).
\end{equation}
After combining the large deviation estimate (\ref{eq:LDTgamma}), the initial scale estimate (\ref{eq:LnL2n}) and the Avalanche Principle (Theorem \ref{thm:AVP}), we obtain the following convergence rate of $L_n(E)$ to $L(E)$: 
\begin{prop}\label{prop:L-Ln}
There exists $N_4\in\N$ explicitly depends on $\tilde{N},\widetilde{N}_0,\Lambda_v,\tilde{c}({v, \rho})$ and $\gamma$. For any $n>N_4$ and 
$(\omega,E)\in \tilde{U} \times I$,
\begin{equation}\label{eq:L-Ln}
 \mid L(E)+L_{n}(E)-2L_{2n}(E) \mid<e^{-\tilde{c}({v, \rho})n\gamma/5}.
\end{equation}
\end{prop}
Proposition \ref{prop:L-Ln} can be derived from an induction method developed by Goldstein and Schlag in \cite{GS1} (see also in \cite{B1},\cite{YZ}). 
For sake of completeness, we include the proof in Appendix \ref{appsec:L-Ln}. 

Another key ingredient for the proof of Theorem \ref{thm:holder1} is the following control on  $\partial_EL_n(\omega,E)$ with respect to $\gamma$. 
\begin{prop}\label{prop:derivLn}
There exists $N_5\in\N$ explicitly depends on $\widetilde{N}_0,\Lambda_v,\tilde{c}({v, \rho})$ and $\gamma$. For any $n>N_5$ and $(\omega,E)\in \tilde{U}\times I$,
\begin{equation}\label{eq:derivLn}
 |\partial_EL_n(E)|\le 2e^{2n\gamma}.
\end{equation}
\end{prop}
Proposition \ref{prop:derivLn} is essentially contained in \cite{B1}, we include the proof in Appendix \ref{appsec:derivLn} with these specific parameters. \\

Now we are in the place to complete the proof of Theorem \ref{thm:holder1} by using (\ref{eq:L-Ln}) and \eqref{eq:derivLn}. 
For short hand we will write $\tilde{c}(c,\rho)$ as $\tilde{c}$, and denote
\beq\label{def:tildec0}
\tilde{c}_0:=\tilde{c}+20.
\eeq

Let $N_6=\max\{N_4,N_5\}$ and 
 \begin{align}\label{def:etaEE'}
 \eta:=\min\left( e^{-2\gamma N_6 \tilde{c}_0 /5},\ 8^{-4\tilde{c}_0/ \tilde{c}} \right)<1.
 \end{align}

Now for any $E,E'\in I$ such that $|E-E'|<\eta$, let 
\begin{equation}\label{eq:nEE'}
n=\lfloor -\frac{5 \log|E-E'|}{\gamma \tilde{c}_0} \rfloor.
\end{equation}
Using the first term in \eqref{def:etaEE'}, it is easy to check that
\begin{align}\label{eq:expnEE'}
\frac{-5 \log|E-E'|}{\gamma \tilde{c}_0}
\geq n
\geq \frac{-5 \log|E-E'|}{2\gamma \tilde{c}_0}
\geq N_6=\max{(N_4, N_5)}.
\end{align}
Now we can apply Proposition \ref{prop:L-Ln} and Proposition \ref{prop:derivLn} to the above $n,E,E'$ to obtain
\begin{eqnarray}
 && |L(E)-L(E')|\nonumber \\
 &\leq  &| L(E)+L_{n}(E)-2L_{2n}(E) |
  +| L(E')+L_{n}(E')-2L_{2n}(E') | \nonumber\\
 && +| L_{n}(E)-L_{n}(E') |+2 | L_{2n}(E)-L_{2n}(E')| \nonumber\\
   &\leq & 2e^{-\tilde{c} n\gamma/5}+ 4e^{2n\gamma}|E-E'|+2e^{4n\gamma}|E-E'| \nonumber\\
   &\leq & 2e^{-\tilde{c} n\gamma/5}+ 6e^{4n\gamma}|E-E'| \label{eq:LE-LE'}.
\end{eqnarray}

In view of the upper and lower bound of $n$ in \eqref{eq:expnEE'}, we have
\begin{equation}\label{eq:ngamma1}
  e^{n\gamma}<|E-E'|^{-5/\tilde{c}_0},
\end{equation}
and
\begin{equation}\label{eq:ngamma2}
  e^{-n\gamma}<|E-E'|^{5/(2\tilde{c}_0)}.
\end{equation}
By (\ref{eq:LE-LE'}),(\ref{eq:ngamma1}) and (\ref{eq:ngamma2}), we have that for all $\omega\in \tilde{U}$, $E,E'\in I$ and $|E-E'|<\eta<1$,
\beq\label{eq:holder}
\begin{aligned}
  |L(E)-L(E')|
\leq &2|E-E'|^{\tilde{c}/(2\tilde{c}_0)}+6|E-E'|^{1-20/\tilde{c}_0} \\
=& 2|E-E'|^{\tilde{c}/(2\tilde{c}_0)}+6|E-E'|^{\tilde{c}/\tilde{c}_0}\\
\leq & 8|E-E'|^{\tilde{c}/(2\tilde{c}_0)}.
\end{aligned}
\eeq
Using the second term in \eqref{def:etaEE'}, we have
\beq\nn
8\leq \eta^{-\tilde{c}/(4\tilde{c}_0)}<|E-E'|^{-\tilde{c}/(4\tilde{c}_0)}.
\eeq
Plugging it into \eqref{eq:holder}, we obtain
\beq\label{eq:EE'}
|L(E)-L(E')| \leq  |E-E'|^{\tilde{c}/(4\tilde{c}_0)}=:|E-E'|^\tau.
\eeq
This proves Theorem \ref{thm:holder1}. \qed 

\subsection{Proof of Theorem \ref{thm:holder2}}
Let $\tilde \lambda$, $b$, $B$ and $N=N(\omega,\lambda,f,\rho)$ be given as in Corollary \ref{cor:LDTlambda}.
 Assume that 
$\lambda>\max\{\tilde \lambda,e^{B\beta(\omega)}\}$, Corollary \ref{cor:LDTlambda} implies that for any $n\geq N$, we have
\begin{align}\label{eq:Omega}
{\rm mes}\left\lbrace x\in\T:\mid u_n(\omega, E; x)-L_n(\omega, E)\mid>\frac{1}{19}\log{\lambda} \right\rbrace
\leq e^{-n\,b\log{\lambda}}.
\end{align}
In view of \eqref{eq:SSlowerbd} and \eqref{eq:D-1}, we have that for $n\geq N$
\begin{align}\label{eq:1819}
\frac{18}{19}\log\lambda\le  L_n(E)\le \frac{20}{19}\log\lambda,\ \  0\le L_n(E)-L_{2n}(E)\le \frac{2}{19}\log\lambda. 
\end{align} 

By (\ref{eq:Omega}), \eqref{eq:1819} and thhe same reasoning for Proposition \ref{prop:L-Ln}, we have
\begin{prop}\label{prop:L-Lnlambda}
Assume that $\beta(\omega)<\infty$ and $\lambda>\max\{\tilde \lambda,e^{B\beta(\omega)}\}$. 
There exists $N_7\in\N$ explicitly depends on $\lambda$ and $b$ such that for any $n>N_7$ and $E\in {\cal N}_{\lambda f}$,
\begin{equation}\label{eq:L-Lnlambda}
 \mid L(E)+L_{n}(E)-2L_{2n}(E) \mid<e^{-\frac{1}{3}n\,b\log \lambda}
\end{equation}
\end{prop}
By the trivial bound $\sup\limits_{n\in\N}\ \sup\limits_{x\in\T} \sup\limits_{E\in{\cal N}_{\lambda f}} u_n(x)\le \Lambda_v\le 2\log \lambda$, we have for any $n,x$ and $E\in {\cal N}_{\lambda f}$, 
$$\Big|\partial_E\log\|M_n(\omega,E;x)\|\Big|\le
  \|\partial_E M_n(\omega,E;x)\| \le\sum_{j=1}^n \|M_{n-j}(x+j\omega;E)\|\cdot\| M_{j-1}(\omega,E;x)\|
  \le ne^{2n\log\lambda},$$
  which implies
\begin{equation}\label{eq:derivlambda}
|\partial_EL_n(\omega,E)|\le e^{2n\log\lambda}. 
\end{equation}

Clearly, by (\ref{eq:L-Lnlambda}), (\ref{eq:derivlambda})  and the same argument from (\ref{def:etaEE'}) to (\ref{eq:EE'}), we can prove \eqref{eq:holder2}. 
More precisely, for all $E,E'\in{\cal N}_{\lambda f}$ satisfying
\begin{align}
|E-E'|<\widetilde\eta:=\min\{e^{-2(12+b) N_7 (\log\lambda) /3}
, 5^{-4(12+b)/b}\},
\end{align}
set $n=\lfloor \frac{3\log|E-E'|^{-1}}{\log\lambda(12+b)} \rfloor$. Then we have
\begin{eqnarray}
  |L(E)-L(E')|&<&2e^{-\frac{1}{3}n\,b\log \lambda}
  + 3e^{4n\log\lambda}|E-E'| \nonumber\\
&\le & 5|E-E'|^{\frac{b}{2(12+b)}}\nonumber\\
&\le&|E-E'|^{\frac{b}{4(12+b)}}=:|E-E'|^{\tilde\tau}. 
\end{eqnarray}
This completes the proof of Theorem \ref{thm:holder2}. \qed

\appendix

\section{Proof of \eqref{eq:GS08uk}}\label{sec:GS08uk}
The proof is essentially contained in \cite[Section II]{BG}, we include a proof here for completeness.
\subsubsection*{Proof of \eqref{eq:GS08uk}}
Let us pick a bump function $\eta(x)$ defined as follows:
\beq\label{def:eta}
\begin{aligned}
\eta(x)=
\begin{cases}
32(x+\frac{3}{4})^3,\qquad\qquad -\frac{3}{4}\leq x<-\frac{1}{2},\\
1-32(x+\frac{1}{4})^3,\qquad\ -\frac{1}{2}\leq x<-\frac{1}{4},\\
1,\qquad\qquad\qquad\qquad\ -\frac{1}{4}\leq x<\frac{1}{4},\\
1-32(x-\frac{1}{4})^3,\qquad\quad\ \frac{1}{4}\leq x<\frac{1}{2},\\
32(x-\frac{3}{4})^3,\qquad\qquad\quad \frac{1}{2}\leq x<\frac{3}{4}.
\end{cases}
\end{aligned}
\eeq
Then it is easy to see that 
\beq\label{eq:eta1}
\begin{aligned}
&{\rm supp}\eta\subset \left[-\frac{3}{4},\frac{3}{4}\right],\ \ \sum_{s\in\Z}\eta(x+s)=1,\ {\rm and \ }\\
&0\leq \eta(x)\leq 1,\ \ |\eta'(x)|\le 6,\ \ |\eta''(x)|\leq 48
\ {\rm for \ all}\ \ x\in\R
\end{aligned}
\eeq

Let $w(x):=\int_{\Omega_1} \log|x-\zeta|\,d\mu(\zeta)$ and $t:=\sup_{\Omega_1}u_n(z)$. Since $u_n(x)$ is 1-periodc on $\R$, we have
\begin{eqnarray}
  \hat u_n(k) &=& \widehat {(u_n-t)}(k) \nonumber \\
   &=&  \int_{-\frac{1}{2}}^{\frac{1}{2}}
(u_n(x)-t)e^{-2\pi {i} kx}\ {\rm d}x  \nonumber \\
   &=&  \int_\R (u_n(x)-t)\eta(x)e^{-2\pi {i} kx}\ {\rm d}x \nonumber\\
&=&  \frac{i}{2\pi k}\int_\R \partial_x\Big((w(x)+h(x)-t)\eta(x)\Big)e^{-2\pi {i} kx}\ {\rm d}x  \nonumber\\
&=&  \frac{i}{2\pi k}\int_\R \partial_x(w\eta)e^{-2\pi {i} kx}\ {\rm d}x + \frac{i}{2\pi k}\int_\R \partial_x\Big((h-t)\eta\Big)e^{-2\pi {i} kx}\ {\rm d}x \nonumber\\
&=&
\frac{i}{2\pi k}\int_\R \eta(x)\partial_xw(x)e^{-2\pi {i} kx}\ {\rm d}x\label{eq:pfukoldw1}\\
&&+
\frac{i}{2\pi k}\int_\R w(x)\partial_x\eta(x)e^{-2\pi {i} kx}\ {\rm d}x\label{eq:pfukoldw2}\\
&&+
\frac{i}{2\pi k}\int_\R\eta(x)\partial_x h(x)e^{-2\pi {i} kx}\ {\rm d}x\label{eq:pfukoldh1}\\
&&+
\frac{i}{2\pi k}\int_\R (h(x)-t)\partial_x\eta(x)e^{-2\pi {i} kx}\ {\rm d}x\label{eq:pfukoldh2}
\end{eqnarray}
Clearly, (\ref{eq:pfukoldh1}),(\ref{eq:pfukoldh2}) can be bounded by
\begin{equation}\label{eq:pfukoldh12}
|(\ref{eq:pfukoldh1})|+|(\ref{eq:pfukoldh2})|
\le \frac{1}{2\pi|k|}\Big(\|\partial_x h\|_{L^\infty(\Omega_0)}
+6\|h-\sup_{\Omega_1}u_n\|_{L^\infty(\Omega_0)}\Big)
\end{equation}

It is enough to estimate (\ref{eq:pfukoldw1}) and (\ref{eq:pfukoldw2}) by (\ref{eq:GS08decomp}). The bound for (\ref{eq:pfukoldw2}) is trivial since
\begin{eqnarray*}
\big|\int_\R w(x)\partial_x\eta(x)e^{-2\pi {i} kx}\ {\rm d}x\big|
&\le& 6 \int_{\Omega_1} \int_{-1}^{1} \Big|\log|x-\zeta|\Big|{\rm d}x  \,d\mu(\zeta) \\
&\le& 6 \int_{\Omega_1} \,d\mu(\zeta) \sup_{\zeta \in \Omega_1}\int_{-1}^1 \Big|\log|x-\zeta|\Big|{\rm d}x  \\
&\le& 6 \mu({\Omega_1} ) \int_{-2}^2 \Big|\log|x|\Big|{\rm d}x\\
&=& (24 \log2) \mu({\Omega_1} )
\end{eqnarray*}

The bound for (\ref{eq:pfukoldw2}) follows from the direct compuation in \cite{BG}:
\begin{eqnarray*}
\big|\int_\R \eta(x)\partial_xw(x)e^{-2\pi {i} kx}\ {\rm d}x\big|
&=&\Big| \int_{\Omega_1} \int_\R
\frac{x-{\rm Re}\zeta}{|x-\zeta|^2}e^{-2\pi {i} kx}\eta(x)
{\rm d}x  \,d\mu(\zeta) \Big|\\
&\le&  \int_{\Omega_1} \Big|\int_\R
\frac{x-{\rm Re}\zeta}{|x-\zeta|^2}e^{-2\pi {i} kx}\eta(x)
{\rm d}x \Big| \,d\mu(\zeta)  \\
&\le& \mu({\Omega_1} )\sup_{\zeta \in \Omega_1}\Big|\int_\R
\frac{x-{\rm Re}\zeta}{|x-\zeta|^2}e^{-2\pi {i} kx}\eta(x)
{\rm d}x \Big| \le C_4  \mu({\Omega_1} ),
\end{eqnarray*}
where $C_4>0$ is some aboslute constant given as in \cite{BG} such that $$\sup_{\zeta \in \Omega_1}\Big|\int_\R
\frac{x-{\rm Re}\zeta}{|x-\zeta|^2}e^{-2\pi {i} kx}\eta(x)
{\rm d}x \Big|\le C_4.$$
This finishes the proof.
\qed

\section{Proof of Lemma \ref{lm:Fourierundelta}}\label{sec:undeltaproof}
Let $\eta(x)$ be the bump function defined as in \eqref{def:eta}.
Then 
\beq\label{eq:undeltak-unk1}
\begin{aligned}
\left|\hat{u}_{n,\delta}(k)-\hat{u}_n(k)\right|
=&\left|\int_{\R}\ \int_{\Omega_1}  \log{\left(\frac{|x-w|+\delta}{|x-w|}\right)}  e^{-2\pi i k x} \eta(x)\ \mu(\mathrm{d} w)\ \mathrm{d} x\right|\\
\leq &\int_{\Omega_1} \left| \int_{\R} \log{\left(\frac{|x-w|+\delta}{|x-w|}\right)}  e^{-2\pi i k x} \eta(x)\ \mathrm{d} x \right|\ \mu(\mathrm{d} w)\\
\leq &\mu(\Omega_1) \sup_{w\in \Omega_1} \left| \int_{\R} \log{\left(\frac{|x-w|+\delta}{|x-w|}\right)}  e^{-2\pi i k x} \eta(x)\ \mathrm{d} x \right|.
\end{aligned}
\eeq
By Lemma \ref{lm:sup-supGS08}, we already have control of $\mu(\Omega_1)$, thus it suffices to estimate the following term for $w=w_1+i w_2\in \Omega_1$:
\beq\label{eq:undeltak-unk2}
\begin{aligned}
&\left| \int_{\R} \log{\left(\frac{|x-w|+\delta}{|x-w|}\right)}  e^{-2\pi i k x} \eta(x)\ \mathrm{d} x \right|\\
= &\left| \int_{-3/4+w_1}^{3/4+w_1} \log{\left(1+\frac{\delta}{\sqrt{x^2+w_2^2}}\right)}  e^{-2\pi i k x} \eta(x+w_1)\ \mathrm{d} x \right|,
\end{aligned}
\eeq
in which we used $\mathrm{supp}(\eta)\subset [-3/4, 3/4]$. Next use the fact that $|\eta(x)|\leq 1$ for any $x\in \R$ and the integrand is monotone decreasing in $x$, we have 
\beq\nn
\begin{aligned}
&\left| \int_{-3/4+w_1}^{3/4+w_1} \log{\left(1+\frac{\delta}{\sqrt{x^2+w_2^2}}\right)}  e^{-2\pi i k x} \eta(x+w_1)\ \mathrm{d} x \right|\\
\leq & \int_{-3/4}^{3/4} \log{\left(1+\frac{\delta}{\sqrt{x^2+w_2^2}}\right)} \ \mathrm{d} x\\
\leq & 2\int_{0}^{3/4} \log{\left(1+\frac{\delta}{x}\right)} \ \mathrm{d} x\\
=& 2\delta \log{\delta^{-1}}+\frac{3}{2}\log\left(1+\frac{4\delta}{3}\right)+2\delta \log\left(\frac{3}{4}+\delta\right).
\end{aligned}
\eeq
Use that $\delta<\frac{1}{144}$, see \eqref{eq:delta<}, and that the following holds,
\begin{align*}
\frac{3}{2}\log\left(1+\frac{4\delta}{3}\right)+2\delta \log\left(\frac{3}{4}+\delta\right)<\delta \log{\delta^{-1}},\ \ \text{for}\ \ 0<\delta<0.15,
\end{align*}
we obtain that
\beq\label{eq:undeltak-unk3}
\left| \int_{-3/4+w_1}^{3/4+w_1} \log{\left(1+\frac{\delta}{\sqrt{x^2+w_2^2}}\right)}  e^{-2\pi i k x} \eta(x+w_1)\ \mathrm{d} x \right|\\
\leq  3 \delta \log{\delta^{-1}}.
\eeq

\eqref{eq:udeltak} follows from combining \eqref{eq:undeltak-unk1}, \eqref{eq:undeltak-unk2} with \eqref{eq:undeltak-unk3}.

The proof of \eqref{eq:udeltak3} follows from a similar idea to that of \eqref{lm:ukold}, the difference is that we need to do integration by parts twice in order to get $(k^2\delta)^{-1}$ Fourier decay.
Let us mention that one needs the control of $\|\partial^2_x h\|_{L^\infty (\Omega_0)}$, which is provided in \eqref{eq:GS08h'}, as well as $|\eta''(x)|\leq 48$ as in \eqref{eq:eta1}.
\qed

\section{Proof of Proposition \ref{prop:L-Ln}}\label{appsec:L-Ln}
\begin{thm}[Avalanche Principle, \cite{GS1}]\label{thm:AVP}
Let $B_1,\cdots,B_m$ be a sequence of unimodular $2\times2$-matrices. Suppose that
\begin{eqnarray}
&&\min_{1\le j\le m}\|B_j\|\ge \mu>m \quad {\rm and} \label{eq:Avp1}\\
&&\max_{1\le j< m}[\log{\|B_{j+1}\|}+\log{\|B_{j}\|}-\log{\|B_{j+1}B_{j}\|}]<\frac{1}{2}\log\mu.\label{eq:Avp2}
\end{eqnarray}
Then
\beq{}
\mid \log{\|B_{m}\cdots B_{1}\|}+\sum_{j=2}^{m-1}\log{\|B_{j}\|}-\sum_{j=1}^{m-1}\log{\|B_{j+1}B_{j}\|}\mid
<C_A\frac{m}{\mu},
\eeq{}
where $C_A$ is an absolute constant. 
\end{thm}

For  any $n\geq N(\omega, \frac{18}{19}\gamma, v, \rho)$ and $E\in I$, set
$$\Omega_n(j)=\{x\in\T:\mid u_n\Big(x+(j-1)n\omega\Big)-L_n(E)\mid>\frac{1}{20}L(E)\}$$
$$\Omega_{2n}(j)=\{x\in\T:\mid u_{2n}\Big(x+(j-1)n\omega\Big)-L_{2n}(E)\mid>\frac{1}{20}L(E) \}$$
$$\Omega=\cup_{j=1}^{m}\Omega_n(j)\bigcup\cup_{j=1}^{m-1}\Omega_{2n}(j),$$
(\ref{eq:LDTgamma}) implies that
$ {\rm mes}\Omega_n(j)\leq e^{-\frac{1}{2}\tilde{c}({v, \rho}) n\gamma }$, ${\rm mes}\Omega_{2n}(j)\leq e^{-\tilde{c}({v, \rho}) n\gamma }$.
Take $m=[n^{-1}\exp({\frac{1}{4}\tilde{c}({v, \rho}) n\gamma })]$ and $n_1=mn$,
then $(2n)^{-1}\exp({\frac{1}{4}\tilde{c}({v, \rho}) n\gamma })<m<n_1<e^{\frac{1}{4}\tilde{c}({v, \rho}) n\gamma }$. Therefore,
\begin{equation}\label{eq:mesOmega}
  {\rm mes}\Omega<2me^{-\frac{1}{2}\tilde{c}({v, \rho}) n\gamma }<2e^{-\frac{1}{4}\tilde{c}({v, \rho}) n\gamma }
\end{equation}
provided $\exp({\frac{1}{4}\tilde{c}({v, \rho}) n\gamma })>2n$.\\

For any $x\not\in\Omega$,
\begin{eqnarray}|u_n\Big(x+(j-1)n\omega\Big)-L_n(E)|<\frac{1}{20}L(E)<\frac{1}{20}L_n(E),\quad j=1,\cdots,m,\label{u-ln}\\
|u_{2n}\Big(x+(j-1)n\omega\Big)-L_{2n}(E)|<\frac{1}{20}L(E)<\frac{1}{20}L_{2n}(E),\quad j=1,\cdots,m-1.\label{u-l2n}
\end{eqnarray}
Thus
\begin{eqnarray}
\frac{19}{20}L_n(E)<u_n(x+(j-1)n\omega)<\frac{21}{20}L_n(E), \label{un}\\
\frac{19}{20}L_{2n}(E)<u_{2n}(x+(j-1)n\omega)<\frac{21}{20}L_{2n}(E). \label{u2n}
\end{eqnarray}
Denote $B_j=M_n(x+(j-1)n\omega)$, then $$u_n(x+(j-1)n\omega)=\frac{1}{n}\log{\|M_n(x+(j-1)n\omega)\|}=\frac{1}{n}\log{\|B_j\|},$$
$$u_{2n}(x+(j-1)n\omega)=\frac{1}{2n}\log{\|M_{2n}(x+(j-1)n\omega)\|}=\frac{1}{2n}\log{\|B_{j+1}B_j\|}.$$
Notice that $\tilde{c}({v, \rho})<1$, by (\ref{un}) and the choice of $m$,
\begin{equation}\label{eq:mu}
  \|B_j\|>e^{\frac{19}{20}nL_n(E)}
>e^{\frac{19}{20}nL(E)}:=\mu>e^{\frac{18}{20} n\gamma}>e^{\frac{1}{4}\tilde{c}({v, \rho}) n\gamma}>m,\quad j=1,\cdots,m.
\end{equation}
By (\ref{eq:LnL2n}), (\ref{u-ln} and \ref{u-l2n}),
\begin{eqnarray}
&&\left |\log{\|B_{j+1}\|}+\log{\|B_{j}\|}-\log{\|B_{j+1}B_{j}\|}\right|  \nonumber\\
&<&\mid \log{\|B_{j+1}\|}-n L_n(E)\mid+\mid\log{\|B_{j}\|}-n L_n(E)\mid   \nonumber\\
   &&+\mid 2n L_n(E)-2n L_{2n}(E)\mid+\mid2n L_{2n}(E)-\log{\|B_{j+1}B_{j}\|}\ \mid    \nonumber\\
&<&\frac{n}{20}L(E)+\frac{n}{20}L(E)+\frac{2n}{20}L(E)+\frac{2n}{20}L(E)     \nonumber\\
&=&\frac{6}{20}nL(E)=\frac{6}{20}\cdot \frac{20}{19}\log\mu<\frac{1}{2}\log\mu.  \label{eq:AvpBj}
\end{eqnarray}
Now (\ref{eq:Avp1}),(\ref{eq:Avp2}) required by Avalanche Principle are full filled.  Apply Theorem \ref{thm:AVP} to $B_j,j=1,\cdots,m,$ we have
$$\mid \log{\|B_{m}\cdots B_{1}\|}+\sum_{j=2}^{m-1}\log{\|B_{j}\|}-\sum_{j=1}^{m-1}\log{\|B_{j+1}B_{j}\|}\mid
<C_A\frac{m}{\mu}.$$
Recall $n_1=mn$, clearly
\begin{eqnarray}
&&\Big| \frac{1}{n_1}\log{\|M_{n_1}(x+(j-1)n\omega)\|}+\frac{1}{m}\sum_{j=2}^{m-1}\frac{1}{n}\log{\|M_n(x+(j-1)n\omega)\|}\nonumber\\
&&-\frac{2}{m}\sum_{j=1}^{m-1}\frac{1}{2n}\log{\|M_{2n}(x+(j-1)n\omega)\|}\Big|
<C_A\frac{m}{n_1\mu}<\frac{C_A}{\mu}.\label{F}
\end{eqnarray}
Denote the sum of the left side of (\ref{F}) by $F(x)$, we have got the above bound of $|F(x)|$ outside the set $\Omega$. For those $x\in\Omega$, we use the upper bound (\ref{trivialbdd}) such that
\begin{equation}\label{eq:F2}
  \sup_{\Omega}|F(x)|<4\Lambda_v
\end{equation}

Integrate $F(x)$ over $\T$, by (\ref{eq:mesOmega}) and (\ref{eq:mu}), for $n>\max\{\widetilde{N}_0(\omega, I, v),N(\omega, \gamma/2, v, \rho)\}$ and $E\in I$, we have
\beq\label{eq:F12}
\begin{aligned}
\left| L_{n_1}(E)+\frac{m-2}{m}L_{n}(E)-\frac{2(m-1)}{m}L_{2n}(E)\right|
=&\left| \int_{\T}F(x){\rm d} x\right|\\
<&\frac{C_A}{\mu}+4\Lambda_v\cdot mes\Omega \\
<&\frac{1}{20}e^{-\frac{1}{5}\tilde{c}({v, \rho})n\gamma}, 
\end{aligned}
\eeq
provided
$$n>\frac{10}{7\tilde{c}({v, \rho})\gamma}\log(40C_A)+\frac{20}{\tilde{c}({v, \rho})\gamma}\log(320\Lambda_v)$$
By (\ref{eq:F12}), (\ref{eq:uniformLn}), (\ref{eq:LnL2n}) and the choice of $m$,
\begin{eqnarray}
\mid L_{n_1}(E)+L_{n}(E)-2L_{2n}(E) \mid
&<&\frac{2}{m}\mid L_{n}(E)-L_{2n}(E) \mid+\frac{1}{20}e^{-\frac{1}{5}\tilde{c}({v, \rho})n\gamma}\nonumber\\
&<&\frac{1}{10}e^{-\frac{1}{5}\tilde{c}({v, \rho})n\gamma} \label{eq:Ln1-Ln0}
\end{eqnarray}
provided
$$\tilde{c}({v, \rho})n\gamma >20\log(80n\gamma).$$
Take $\tilde{n}=2n_1=2mn$, the above argument also shows that
\begin{equation}\label{eq:L2n1-Ln0}
  \mid L_{2n_1}(E)+L_{n}(E)-2L_{2n}(E) \mid<\frac{1}{10}e^{-\frac{1}{5}\tilde{c}({v, \rho})n\gamma} .
\end{equation}
Therefore,
\begin{equation}\label{eq:L2n1-Ln1}
  \mid L_{2n_1}(E)-L_{n_1}(E)\mid<\frac{2}{10}e^{-\frac{1}{5}\tilde{c}({v, \rho})n\gamma} < \frac{1}{40}\gamma <\frac{1}{20}L(E),
\end{equation}
provided $n>5(\tilde{c}({v, \rho})\gamma)^{-1}\log(8\gamma^{-1})$.\\

Let $n_0=n$ and for $s=0,1,\cdots$, let 
\begin{align}\label{def:ns}
n_{s+1}=n_s[n_s^{-1}e^{\frac{1}{4}\tilde{c}({v, \rho}) n_s\gamma }]. 
\end{align}
 Inductively, we can prove that
\begin{prop}[Iteration of $L_n(E)$]\label{iteration}
\noindent
\begin{description}
\item[$1^{s}$]
\begin{eqnarray}\mid L_{n_{s+1}}(E)+L_{n_s}(E)-2L_{2n_s}(E) \mid<\frac{1}{10}e^{-\frac{1}{5}\tilde{c}({v, \rho})n_s\gamma},\nonumber\\
\mid L_{2n_{s+1}}(E)+L_{n_s}(E)-2L_{2n_s}(E) \mid<\frac{1}{10}e^{-\frac{1}{5}\tilde{c}({v, \rho})n_s\gamma}. \label{1s}
\end{eqnarray}
\item[$2^{s}$]
\beq{}\label{2s}
\mid L_{2n_{s+1}}(E)-L_{n_{s+1}}(E)\mid<\frac{2}{10}e^{-\frac{1}{5}\tilde{c}({v, \rho})n_s\gamma} < \frac{1}{40}\gamma <\frac{1}{20}L(E)
\eeq{}
\item[$3^{s}$]
\beq{}\label{3s}
\mid L_{n_{s+1}}(E)-L_{n_{s}}(E)\mid<\frac{1}{2}e^{-\frac{1}{5}\tilde{c}({v, \rho})n_{s-1}\gamma}, \quad n_0=n.
\eeq{}
\end{description}
\end{prop}

 Once we have $1^{s-1},2^{s-1}$, we prove $1^{s}$ first as (\ref{eq:Ln1-Ln0}),(\ref{eq:L2n1-Ln0}). Then $2^{s}$ directly follows from $1^{s}$ as (\ref{eq:L2n1-Ln1}). By $1^{s}$ and $2^{s-1}$, we get $3^{s}$ as follows:
\begin{eqnarray*}
&&\mid L_{n_{s+1}}(E)-L_{n_{s}}(E)\mid\\
&<&\mid L_{n_{s+1}}(E)+L_{n_s}(E)-2L_{2n_s}(E) \mid+2\mid L_{n_{s}}(E)-L_{2n_{s}}(E)\mid\\
&<&\frac{1}{10}e^{-\frac{1}{5}\tilde{c}({v, \rho})n_s\gamma}+\frac{4}{10}e^{-\frac{1}{5}\tilde{c}({v, \rho})n_{s-1}\gamma}\\
&<&\frac{1}{2}e^{-\frac{1}{5}\tilde{c}({v, \rho})n_{s-1}\gamma}. \hspace{6cm} \Box
\end{eqnarray*}

When the iteration is established for all $s\ge1$, it is easy to check $n_{s-1}>sn$ by (\ref{def:ns}), we have then
\begin{eqnarray}
\mid L(E)-L_{n_1}(E)\mid&\le&\sum_{s=1}^{\infty}\mid L_{n_{s+1}}(E)-L_{n_{s}}(E)\mid\nonumber\\
                        &\le&\frac{1}{2}\sum_{s=1}^{\infty}e^{-\frac{1}{5}\tilde{c}({v, \rho})n_{s-1}\gamma} \nonumber\\
                        &\le&\frac{1}{2}\,\frac{e^{-\frac{1}{5}\tilde{c}({v, \rho})n\gamma}}{1-e^{-\frac{1}{5}\tilde{c}({v, \rho})n\gamma}} \nonumber\\
                        &\le&\frac{9}{10}e^{-\frac{1}{5}\tilde{c}({v, \rho})n\gamma}, \label{eq:L-Ln1}
\end{eqnarray}
provided $e^{-\frac{1}{5}\tilde{c}({v, \rho})n\gamma}<\frac{4}{9}$. 
 
By (\ref{eq:Ln1-Ln0}), we have
\begin{equation}\label{eq:L-Ln0}
  \mid L(E)+L_{n}(E)-2L_{2n}(E) \mid<e^{-\frac{1}{5}\tilde{c}({v, \rho})n\gamma}
\end{equation}

\section{Proof Proposition \ref{prop:derivLn}}\label{appsec:derivLn}
It is enough to show that for $n$ large
\begin{equation}\label{eq:derivMn}
  \sup_{x\in\T} \Big|\partial_E\log\|M_n(\omega,E;x)\|\Big|\le 2ne^{2n\gamma}
\end{equation}
Lemma \ref{lm:uniformtildeN0} and (\ref{eq:lypgamma}) imply that for $n>\tilde N_0$, for any $x\in\T$ and $E\in I$
\begin{equation}\label{eq:uniformUNgamma}
  u_n(\omega,E;x)\le 2\gamma
\end{equation}
i.e., $\|M_j(\omega,E;x)\|\le e^{2n\gamma}$ for $j>\tilde N_0$. For $j\le \tilde N_0$, we use the trivial bound
\begin{equation}\label{eq:Mj2}
  \|M_j(\omega,E;x)\|\le e^{j\Lambda_v}\le e^{\tilde N_0\Lambda_v}:=C_5
\end{equation}
Direct computation shows that for any $x\in\T$, and $n>2C_5\tilde N_0>2\tilde N_0$,
\begin{align*}
\Big|\partial_E\log\|M_n(\omega,E;x)\|\Big|
\le &\|\partial_E M_n(\omega,E;x)\| \\
 \le&\sum_{j=1}^n \|M_{n-j}(x+j\omega;E)\|\cdot\| M_{j-1}(\omega,E;x)\| \\
   =& \sum_{j=1}^{\tilde N_0} + \sum_{j=\tilde N_0+1}^{n-\tilde N_0}+ \sum_{j=n-\tilde N_0+1}^{ n} \\
 \le& \sum_{j=1}^{\tilde N_0} C_5e^{2(n-j)\gamma} +
\sum_{j=\tilde N_0+1}^{n-\tilde N_0}e^{2(n-j)\gamma}\cdot e^{2(j-1)\gamma}
+ \sum_{j=n-\tilde N_0+1}^{ n} C_5e^{2(j-1)\gamma} \\
    \le& 2ne^{2n\gamma}.
\end{align*}

\section{Proofs of (\ref{eq:FRkleq}),(\ref{eq:lowqsR}),(\ref{eq:low1+qsR})}\label{sec:proofofFRk}
\noindent\textbf{ Proof of \eqref{eq:FRkleq}}:

First, trivially we have $F_R(k)\leq 1$.
Direct computation shows:
\begin{eqnarray*}
0\leq F_R(k)
=\frac{\sin^2{(\pi R k\omega})}{R^2\sin^2{(\pi k\omega})}
=\frac{\sin^2{(\pi R \|k\omega\|_{\T}})}{R^2\sin^2{(\pi \|k\omega}\|_{\T})}
\le\frac{\sin^2{(\pi R \|k\omega\|_{\T}})}{4R^2\|k\omega\|^2},
\end{eqnarray*}
in which we used $\sin{(\pi x)}\geq 2x$ for $0\leq x\leq 1/2$.

Distinguishing the cases $R\|k\omega\|_{\T}\geq 1$ and $R\|k\omega\|_{\T}<1$, one can easily prove the stated bound.\qed

\noindent {\bf Proofs of (\ref{eq:lowqsR}) and (\ref{eq:low1+qsR}):}

Since $\frac{p}{q}$ is a continued fraction approximant of $\omega$, we have $|\omega-\frac{p}{q}|<\frac{1}{q^2}$. 
This implies that for any $0\neq |k|<\frac{q}{2}$, $\left|k\omega-\frac{kp}{q}\right|<\frac{k}{q^2}<\frac{1}{2q}$, and hence 
\beq\label{eq:komega>1/2q}
\|k\omega\|_{\T}\ge \|kp/q\|_{\T}-\left|k\omega-\frac{kp}{q}\right|\geq \frac{1}{2q}.
\eeq 

If we take $j_1\neq j_2\in(0,\frac{q}{4}]\in \Z$, then clearly $|j_1\pm j_2|<\frac{1}{2q}$. 
Thus by \eqref{eq:komega>1/2q}, 
$\Big|\|j_1\omega\|_{\T}-\|j_2\omega\|_{\T}\Big|\geq \min{(\|(j_1+j_2)\omega\|_{\T}, \|(j_1+j_2)\omega\|_{\T})}\geq\frac{1}{2q}$. 
This implies that $\{\|k \omega\|_{\T}\}_{k=1}^{[\frac{q}{4}]}$ are $\frac{1}{2q}$ departed, and by \eqref{eq:komega>1/2q} the smallest one is $\geq \frac{1}{2q}$. 
If we rearrange them in the increasing order and label them as $\|k_{1}\omega\|_{\T}<\|k_{2}\omega\|_{\T}<\cdots<\|k_{[q/4]}\omega\|_{\T}$,
then $\|k_{s}\omega\|_{\T}\ge\frac{s}{2q}.$
Hence
$$\sum_{1\le |k|<\frac{q}{4}}\frac{1}{1+R^2\|k\omega\|_{\T}^2}
=2\sum_{1\leq k<\frac{q}{4}}\frac{1}{1+R^2\|k\omega\|_{\T}^2}
\le 2\sum_{s=1}^{[q/4]}\frac{1}{1+R^2{(\frac{s}{2q})}^2}
\leq \frac{4q}{R}\int_0^{\infty} \frac{\rm{d}x}{1+x^2}
=2\pi\frac{q}{R},$$
this proved \eqref{eq:lowqsR}.

For $\ell \geq 1$, let $I_{\ell}:=[\frac{q}{4}\ell,\frac{q}{4}(\ell+1))\cap \Z, \ell\ge1$. 
We divide $I_{\ell}$ into two disjoint sets, $S_1=\{k\in I_{\ell},|k\omega-[k\omega]|<0.5\}$, 
$S_2=\{k\in I_{\ell},|k\omega-[k\omega]|>0.5\}$. 
Then for $j_1\neq j_2\in I_{\ell}$ belonging to the same subset (either $S_1$ or $S_2$), 
we have $\Big|\|j_1\omega\|_{\T}-\|j_2\omega\|_{\T}\Big|=\|(j_1- j_2)\omega\|_{\T}$.
Since clearly $|j_1- j_2|<\frac{q}{4}$, by \eqref{eq:komega>1/2q}, we have $\|(j_1- j_2)\omega\|_{\T}\geq \frac{1}{2q}$.
This implies that $\{\|k\omega\|_{\T}\}_{k\in S_1}$ are $\frac{1}{2q}$ apart from each other, and the same holds for $S_2$.
Thus we could arrange the terms $\{\|k\omega\|_{\T}\}_{k\in S_1\ \text{(or } S_2 \text{)}}$ in the increasing order and label them as 
$\|k_1\omega\|_{\T}<\|k_2\omega\|_{\T}<\cdots \|k_{[q/4]}\omega\|_{\T}$, and we have $\|k_s\omega\|_{\T}\geq \frac{s-1}{2q}$.
Hence
\begin{align*}
\sum_{|k|\in[\frac{q}{4}l,\frac{q}{4}(l+1))}\frac{1}{1+R^2\|k\omega\|_{\T}^2}
=&2\sum_{k\in I_{\ell}}\frac{1}{1+R^2\|k\omega\|_{\T}^2}
=2\left(\sum_{k\in S_1}+\sum_{k\in S_2}\right)\frac{1}{1+R^2\|k\omega\|_{\T}^2}\\
\leq &2\sum_{s=1}^{[q/4]}\frac{1}{1+R^2{(\frac{s-1}{2q})}^2}
\le 2+2\frac{4q}{R}\int_{0}^{\infty}\frac{\rm{d}x}{1+x^2}=2+ 4\pi\frac{q}{R},
\end{align*}
this proves \eqref{eq:low1+qsR}.


\vspace{1cm}
Rui Han, 

School of Math, Institute for Advanced Study.

Current address:

School of Math, Georgia Institute of Technology

E-mail address:  rui.han@math.gatech.edu

\vspace{1cm}
Shiwen Zhang, 

Dept. of Math., Michigan State University. 

Current  address: 

School of Math, University of Minnesota

E-mail address: zhan7294@umn.edu

\end{document}